\newcommand{\be}{\begin{equation}}
\newcommand{\ee}{\end{equation}}
\newcommand{\bea}{\begin{eqnarray}}
\newcommand{\eea}{\end{eqnarray}}
\newcommand{\ba}[1]{\begin{array}{#1}}
\newcommand{\ea}{\end{array}}
\newcommand{\Ket}[1]{|#1\rangle}
\newcommand{\Bra}[1]{\langle#1|}

\documentclass[pra,aps,showpacs,twocolumn]{revtex4-1}
\usepackage{times}
\usepackage{amssymb}
\usepackage{amsmath}
\usepackage{mathrsfs}
\usepackage{graphicx}
\usepackage{epsfig}
\usepackage{dcolumn}
\usepackage{color}
\usepackage{bm}

\begin{document}

\title{A wave-function ansatz method for calculating field correlations and its application to the study of spectral filtering and quantum dynamics of multi-emitter systems}
\author{Sumanta Das$^{1}$, Liang Zhai $^{2,3}$, Mantas \v{C}epulskovskis$^{2}$, Alisa Javadi$^{2,3}$, Sahand Mahmoodian$^{2,5}$, Peter Lodahl$^{1}$ and Anders S. S\o rensen$^{1}$}
\affiliation{$^1$Center for Hybrid Quantum Networks (Hy-Q), Niels Bohr Institute, University of Copenhagen, Blegdamsvej 17, DK-2100 Copenhagen \O, Denmark\\
$^2$ Niels Bohr Institute, University of Copenhagen, Blegdamsvej 17, 2100 Copenhagen \O, Denmark\\
$^3$Department of Physics, University of Basel, Klingelbergstrasse 82 CH 4056 Basel, Switzerland\\
$^5$Institute for Theoretical Physics, Institute for Gravitational Physics (Albert Einstein Institute), Leibniz University Hannover, Appelstra$\beta$e 2, 30167 Hannover, Germany}
\date{\today}
\begin{abstract}
We develop a formalism based on a time-dependent wave-function ansatz to study correlations of photons emitted from a collection of two-level quantum emitters. We show how to simulate the system dynamics and evaluate the intensity of the scattered photons and the second-order correlation function $g^{(2)}$ in terms of the amplitudes of the different components of the wave function. Our approach is efficient for considering systems that contain up to two excitations. To demonstrate this we first consider the example of spectral filtering of photons emitted from a single quantum emitter. We show how our formalism can be used to study spectral filtering of the two-photon component of the emitted light from a single quantum emitter for various kinds of filters. Furthermore, as a general application of our formalism, we show how it can be used to study photon-photon correlations in an optically dense ensemble of two-level quantum emitters. In particular we lay out the details of simulating correlated photon transport in such ensembles reported recently by S. Mahmoodian {\it et.al.} [Phys. Rev. Lett. {\bf 121}, 143601 (2018)].  Compared to other existing techniques, the advantage of our formalism is that it is applicable to any generic spectral filter and quantum many-body systems involving a large number of quantum emitters while requiring only a modest computational resource.
\end{abstract}

\pacs{} 
\maketitle

\section{Introduction}
Understanding the interaction of light with quantum emitters is one of the fundamental question in quantum optics \cite{Chang_14, Reiserer_15, Diaz_19}. The importance of this extremely rudimentary process lies in the fact that it is the basis of a wide variety of physical processes ranging from natural phenomenon like photosynthesis to advanced technologies like digital imaging \cite{Herek_02, Tenne_19}. Furthermore, it is also central to the development of modern disruptive technologies like quantum computing \cite{Brien09, Guzik_12} and quantum communications \cite{Kim08, Sangouard_12,Diamanti_16, Mattar_18}. Several methods have been developed over the years to study light-matter interaction at different regimes and in different systems \cite{Kojima03, Shen07, Yud08, With10, Zheng10, Liao10, Fan10, Plet12, Ring14, Shi15, Garcia17, Man17, Das118, Das218}. In particular, methods to investigate photon correlations \cite{Zheng10, Anders07, Firstenberg13}, typically involve a density matrix description and exploit the quantum regression theorem \cite{Scullyb, Ficekb}. However, these techniques often become extremely demanding even numerically, as the Hilbert space of the system expands and thus warrants new methods to address light-matter coupling for large systems. 

The existing theoretical approaches have drawbacks for studying even multi-photon emission processes from a single quantum emitter. For example, consider the problem of spectral filtering of emitted photons from a typical single photon source- a single quantum emitter excited by short intense optical pulses. These intense laser pulses contain many photons and can excite the quantum emitter more than once within a single pulse duration reducing the purity of the single photon source \cite{Lukas17}. This problem can be ameliorated using a spectral filter that preferentially removes the two photon component in emission. The theoretical description of the filtering process can be quite cumbersome since one needs to study the spectrum of both single-photon and two-photon emission. Single photon sources \cite{Aharonovich_11, Mizouchi_12,Yuan_02, Strauf_07, Lodahl_15, Ding16, Aharonovich_16, Gabi17, Hennrich_04, Maurer_04} are typically characterized by calculating the single-photon and two time two-photon correlation functions of the form $G^{(1)}(t,t')=\langle :\hat E^\dagger(t)\hat E(t'):\rangle$ and $G^{(2)}(t',t)=\langle : \hat E^\dagger(t)\hat E^\dagger(t')\hat E(t')\hat E(t):\rangle$ respectively, where $\hat E(t) (\hat E(t)^\dagger)$ is a photon annihilation (creation) operator removing (creating) a photon arriving at the detector at time $t$. A standard approach for calculating this correlation function is to use the quantum regression theorem on the density matrix \cite{Ficekb}. However, in the presence of a frequency filter, we need to calculate the correlations as a function of frequencies, for example $G^{(2)}(\omega_{1},\omega_{2})=\langle : \hat{I}(\omega_{1})\hat{I}(\omega_2): \rangle$, where $\hat{I}(\omega) \sim \int dt \int dt' e^{-i\omega (t-t')}\hat E^\dagger(t)\hat E(t')$ is the intensity operator for the frequency $\omega$ and $\langle:\quad:\rangle$ stands for a normally ordered expectation value. This requires calculating a four point correlation function $\tilde{G}^{(2)}(t_1,t_2,t_3,t_4)=\langle :\hat E^\dagger(t_1)\hat E^\dagger(t_2)\hat E(t_3)\hat E(t_4):\rangle$ using the quantum regression theorem. This makes the problem quite intractable to solve as it would require evaluating $O(n^4)$ quantities, where $n$ is the number of grid points in time. Furthermore, the quantum regression method is also unsuitable for evaluating the correlation functions for multi-emitter systems, a common example being multiple emitters in waveguides which are currently under intense investigation \cite{Lodahl_15, Sip16, Javadi18, Tomaso19, Corzo19}. For $N$-emitters and allowing for a total of $m$ excitations in the system, the Hilbert space is $\sim N^{m}$-dimensional and the size of the density matrix is quadratic in the size of the Hilbert space $O(N^{2m})$. Hence the simulation of the density matrix is quite difficult for large $N$ even with modest $m$.

To address these problems, we introduce a wave function ansatz method written in the Schr\"odinger picture for calculating field correlations of photons emitted by a single quantum emitter under pulsed excitation \cite{Liang_thesis} and an ensemble of driven two-level quantum emitters \cite{Mantas_thesis}. Within this paradigm, our formalism can incorporate several degree of freedoms, like coupling of multiple quantum emitters to a $1$D waveguide with an arbitrary placement, chiral and non-chiral waveguide coupling, and arbitrary strength of the coupling as long as the interaction remain Markovian in nature \cite{Sahand18}. We consider a weak coherent state as an input field, incident from the sides or through the $1$D waveguide. We consider the limit in which the emitters can emit a limited number of photons and, achieve simple equations of motion by truncating the wave-function to accommodate only a finite number of emissions (restricted here to two). This is the natural situation for single photon sources which are suppose to emit only a few photons but, is, e.g., also applicable to weakly driven multi-emitter systems. Considering an explicit example of a single quantum emitter coupled to a $1$D waveguide, we show how our formalism can be used to study spectral filtering of two-photon emission. Such a filtering process can be used to enhance the purity of a single photon source \cite{Lukas17}. Additonally, in a recent work we have analytically predicted strongly correlated photon transport in an optically dense ensembles of two level quantum emitters (atoms) with completely chiral coupling and used the present technique to simulate non-perfect chirality \cite{Sahand18}. In this article we present the details of this simulation. 

In comparison to the quantum regression theorem, which requires calculating $O(n^4)$ numbers, our wave function ansatz method only requires calculating $O(n^2)$ numbers in the two-photon component to evaluate the effect of spectral filtering. Our wave function ansatz method closely resembles the well known Monte Carlo wave function technique \cite{Dum92, Molmer93, Plenio98} where the evolution is governed by a non-Hermitian Hamiltonian interrupted by quantum jumps at random times. Indeed our derived equations of motions in section III are identical to the equations for the no-jump evolution of the Monte Carlo wave function for the considered system. As opposed to the random quantum jumps of the Monte Carlo approach however, we consider a full wave-function ansatz of the entire system and reservoir such that the resultant state is a superposition of all possible emission times. The Monte Carlo wave function approach was originally developed to avoid the quadratic increase in the size of the problem from going to the density matrix and thus shares the same advantage over the quantum regression theorem for large Hilbert spaces as discussed above. However, our method has the advantage that it gives an explicit description of the outgoing quantum state. This can be a conceptual advantage in problems where the outgoing photons are not just detected after leaving the system, but are subject to further evolution. An example of this are quantum information protocols where an emitted photon is part of a subsequent quantum evolution \cite{Sahand16, Das17, Lodahl17}. Here the access to the full state of the outgoing photons makes it simpler to describe the subsequent evolution. Another example where the present approach is advantageous is the study of two-photon correlations from spectrally filtered photon source. With standard Monte Carlo methods such spectral filtering processes can be evaluated only for a Lorentzian filter \cite{Atac94}, whereas our approach is more general and applies to any filter function. This can potentially be useful for improving, e.g. quantum dot single photon sources \cite{Lodahl_15}.

The article is organized as follows: In Sec. II we introduce the model system of  $N$ two level quantum emitters coupled to a $1$D waveguide and discuss the relevant Hamiltonian. In Sec. III. A. we introduce the time-dependent wave-function ansatz for the system and derive the equations of motion for a single two level emitter in III. B. In Sec. III. C. we evaluate the normalized photon-photon correlation function $g^{(2)}$ for a two-level emitter coupled to the $1$D waveguide. In Sec. IV. A. we use our formalism to study the correlation characteristics of photons emitted by a quantum emitter driven by pulsed excitations. In Sec. IV. B. we then apply our  formalism to investigate the method of spectral filtering of two-photon components and its effect on the two-photon correlation $g^{(2)}$. In Sec. V.A. we discuss the steady state behavior of the photon-photon correlation function for emission from a single quantum emitter coupled to a $1$D waveguide. In Sec. V. B. we generalize our formalism to study quantum many-body systems. In particular we give a similar discussion of steady state $g^{(2)}$ but for $N$-two level emitters coupled to a non-chiral $1$D waveguide.  Finally, in Sec. VI we summarize our results. Several details of our calculations are relegated to the appendices.  In appendix A, we provide details of the equations of motion for single two-level emitter coupled to a $1$D waveguide in dimensionless units. In appendix B, we give details of the equation of motion in dimensionless units for the $N$ two-level emitter system. In appendix C, we give detail derivation of the single and two photon correlation function.
\section{Model System and Hamiltonian}
\begin{figure}[t!]
	\includegraphics[height = 3.3 cm]{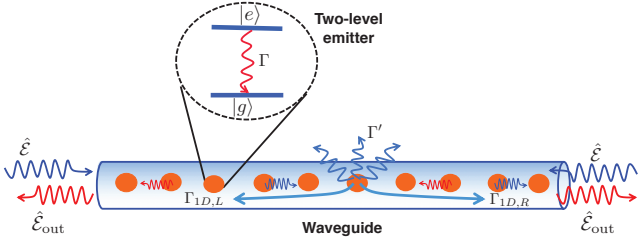}
	\caption{Schematic of multiple two-level quantum emitters coupled to the field mode of a $1$D waveguide and interacting with the incoming field. Here $\Gamma$ is the natural line-width of the two level emitters, $\Gamma^{'}$ is the decay out of the waveguide while $\Gamma_{1D, R/L}$ is the decay of the emitter into the $1D$ waveguide along the right and left, respectively. The excitation and emission fields are represented by the operators $\hat{\mathcal{E}}$ and $\hat{\mathcal{E}}_{\text{out}}$ respectively. \label{fig1}}
\end{figure}

In this section we introduce our theoretical model and the Hamiltonian that governs the dynamics of the systems. We consider a chain of quantum emitters spatially located along a $1$D optical waveguide as shown schematically in Fig. \ref{fig1}. Although we consider here a one dimensional waveguide, with a suitable redefinition of the optical modes, the model can be applied equally to, e.g., a single emitter in free space. Describing it in a waveguide allows for a convenient notation and also allows us to describe how the formalism is applicable for the experimentally relevant system of multiple emitters coupled to a waveguide  \cite{Goban14, Arcari14, Mitsch14, Sahand18}. We assume each of the emitters to be a two-level system with a dipole transition $|e_{j}\rangle \leftrightarrow |g_{j}\rangle$ coupled to the field mode in the $1$D waveguide, where $|e_{j}\rangle$ and $|g_{j}\rangle$ are the excited and ground state respectively of the $j^{th}$ emitter. Note that our approach is generic and can easily be generalized to a wide variety of multi-level emitters like atoms, molecules, quantum dots, superconducting qubits and nitrogen vacancies. 

We next consider the interaction of the emitters with a waveguide mode. The Hamiltonian of our model system is then given by $\hat{\mathcal{H}} = \hat{\mathcal{H}}_{F}+\hat{\mathcal{H}}_{e}+\hat{\mathcal{H}}_{I}$, where the free field Hamiltonian of the multimode electromagnetic field is given by $\hat{\mathcal{H}}_{F} = \int~dk~\hbar\omega_{k}~a^\dagger_k a_k$, while $\hat{\mathcal{H}}_{e}$ is the free energy Hamiltonian of the emitters given by $\sum_{j}\hbar\omega^{j}_{eg}\hat{\sigma}^{j}_{ee}$. Here $a_{k} (a^\dagger_{k})$ is the bosonic field mode annihilation (creation) operator with frequency $\omega_{k}$ while $\omega_{eg}$ is the transition frequency between the excited and ground state. The total interaction Hamiltonian for our model system is given by $\hat{\mathcal{H}}_{I} = \sum_{j}\hat{\mathcal{H}}^{j}_{I}$, where $\hat{\mathcal{H}}^{j}_{I}$ is the interaction Hamiltonian describing the coupling of the field with an emitter located at a spatial position $\text{z}_{j}$ along the waveguide and is given by \cite{Chang07}
\bea
\label{eq1}
\hat{\mathcal{H}}^{j}_{I} = -\hbar \int~dk~\mathcal{G}_{jk}\hat{\sigma}^{+}_{j}\hat{a}_{k}e^{-i(\omega_{k}t-k\text{z}_{j})}+ H.c.
\eea
Here $\mathcal{G}_{jk}$ is the coupling strength between the $j^{th}$ emitter and the field, $\hat{\sigma}_{j} (\hat{\sigma}^{+}_{j})$ are the atomic lowering (raising) operators defined by $\hat{\sigma}_{j} = |g_{j}\rangle\langle e_{j}|$ with $\hat{\sigma}^{+} = [\hat{\sigma}]^{\dagger}$ and satisfying the standard angular momentum commutation relation while $\omega_{k}$ is the frequency corresponding to the propagation wave vector $k$. 

The Hamiltonian introduced above is defined in the Schr\"{o}dinger picture. However, it is more convenient to study the dynamics in the interaction picture. As such we use the unitary operator $\hat{\mathcal{U}} = e^{-i(\hat{\mathcal{H}}_{f}+\hat{\mathcal{H}}_{e})t}$ to transform the Hamiltonian $\hat{\mathcal{H}} = \sum_{j}\hat{\mathcal{H}}_{j}$ from the Schr\"{o}dinger to the interaction picture yielding,
\bea
\label{eq2}
\hat{\mathcal{H}}_{j} = -\hbar\int~dk~\mathcal{G}_{jk}\hat{\sigma}^{+}_{j}\hat{a}_{k}e^{-i\Delta_{kj}t+ik\text{z}_{j}} + H.c.
\eea
Here $\Delta_{kj} = \omega_{k} - \omega^{j}_{eg}$ is the detuning of the $j^{th}$ emitter with respect to the frequency of the incoming field. For generality, we consider the waveguide to support both left- and right-propagating modes. As such the interaction Hamiltonian in Eq. (\ref{eq2}) includes the interaction with both these modes. Furthermore, we assume that the emitter can also couple to the vacuum reservoir modes outside of waveguide. This coupling to reservoir modes model the loss of photons out of the waveguide. To include all these couplings, we then follow \cite{Chang07}, and write the mode operator in the linear dispersive regime as a sum of a right $(\mathrm{R})$, left $(\mathrm{L})$ and side $(\mathrm{S})$ going modes in the form $\mathcal{G}_{jk}\hat{a}_{k}\mathrm{e}^{ik\text{z}_j}\rightarrow(\mathcal{G}_{\text{R}jk}\hat{a}_{\mathrm{R},k}\mathrm{e}^{ik\text{z}_j}+\mathcal{G}_{\text{L}jk}\hat{a}_{\mathrm{L},k}\mathrm{e}^{-ik\text{z}_j} +\mathcal{G}_{\text{S}jk}\hat{a}_{\mathrm{S},k}\mathrm{e}^{ik\text{y}_j})$. Here each mode represents independent quantum fields. The Hamiltonian in Eq. (\ref{eq2}) can then be written as \cite{Shen09}
\bea
\label{eq3}
\hat{\mathcal{H}}_{j} & = &  -\hbar\int d k~\hat{\sigma}^{+}_{j}\bigg(\mathcal{G}_{\mathrm{R}jk}\hat{a}_{\mathrm{R},k}e^{-i\Delta_{kj} t+ik\text{z}_j}+\mathcal{G}_{\mathrm{S}jk}\hat{a}_{\mathrm{S},k}\nonumber\\
&\times&e^{-i\Delta_{kj} t+ik\text{y}_j}+\mathcal{G}_{\mathrm{L}jk}\hat{a}_{\mathrm{L},k}e^{-i\Delta_{kj} t-ik\text{z}_j}\bigg)+H.c., 
\eea
where $\mathcal{G}_{\mathrm{L}(\mathrm{R})jk}$ and $\mathcal{G}_{\text{S}jk}$ are the emitter-field coupling strengths to the left (right) propagating photons in the waveguide and to the outside vacuum reservoir respectively . 

We next introduce a set of slowly-varying field mode operators \cite{Mantas_thesis}, 
\bea
\label{eq4}
\hat{E}_{\mathrm{R}}(\text{z})&=&\frac{1}{\sqrt{2\pi}}\int\mathrm{d}k~\hat{a}_{\mathrm{R},k}\mathrm{e}^{i[(k-k_{0})\text{z}-(\omega_{k}-\omega_{0})t]},\nonumber\\
&=&\frac{1}{\sqrt{2\pi}}\int\mathrm{d}k~\hat{a}_{\mathrm{R},k}\mathrm{e}^{i(k-k_{0})(\text{z}-v_{g}t)}\\
\label{eq4a}
\hat{E}_{\mathrm{L}}(\text{z})&=&\frac{1}{\sqrt{2\pi}}\int\mathrm{d}k~\hat{a}_{\mathrm{L},k}\mathrm{e}^{-i[(k-k_{0})\text{z}-(\omega_{k}-\omega_{0})t]},\nonumber\\
&=&\frac{1}{\sqrt{2\pi}}\int\mathrm{d}k~\hat{a}_{\mathrm{L},k}\mathrm{e}^{-i(k-k_{0})(\text{z}-v_{g}t)},\\
\hat{E}_{\mathrm{S}j}(\text{y})&=&\frac{1}{\sqrt{2\pi}}\int\mathrm{d}k~\hat{a}_{\mathrm{S},k}\mathrm{e}^{ik\text{y}}
\label{eq4b}
\eea
which are functions of the spatial variable $\text{z}$ in the waveguide and the co-ordinate to the side $\text{y}$. Note that in Eq. (\ref{eq4}) and  (\ref{eq4a}) we have expanded $\omega_{k}$ about the central frequency $\omega_{0}$ and defined $v_{g}$ as the group velocity in the medium with $k_{0}$ being the wavevector corresponding to the central frequency $\omega_{0}$ of the photons travelling in the waveguide. The spatially dependent field operators in the above equations satisfy standard commutation relations $\left[\hat{E}_{\mathrm{j}}(\text{z}),\hat{E}_{\mathrm{j}}^{\dagger}(\text{z}')\right]=\delta(\text{z}-\text{z}')$ and $\left[\hat{E}_{\mathrm{j}}(\text{z}),\hat{E}_{\mathrm{k}}^{\dagger}(\text{z})\right]=0$ for $j \neq k$, for each of the $\{j,k \}= \{\text{R},\text{L}\}$ respectively. Furthermore, the operators for the field going to the side satisfy the commutation relation $\left[\hat{E}_{\mathrm{S}j}(\text{y}),\hat{E}_{\mathrm{S}k}^{\dagger}(\text{y}')\right]=\delta_{jk}\delta(y-y')$ and we have assumed independent reservoir for each emitters. On substituting Eq. (\ref{eq4}) - Eq. (\ref{eq4b}) into Eq.(\ref{eq3}), we get a spatially dependent Hamiltonian of the form 
\bea
\label{eq5}
&&\hat{\mathcal{H}}_{j} = -\hbar\sqrt{2\mathrm{\pi}}\bigg[\int d\text{z}~\delta(\text{z}-\text{z}_{j})~\hat{\sigma}^{+}_{j}\bigg(\mathcal{G}_{\mathrm{R}j}\hat{E}_{\mathrm{R}}(\text{z})e^{-i(\Delta t-k_{0}\text{z})}\nonumber\\
&& +\mathcal{G}_{\mathrm{L}j}\hat{E}_{\mathrm{L}}(\text{z})~e^{-i(\Delta t+k_{0}\text{z})}\bigg)+\int ~dy ~\delta(y)\mathcal{G}_{\mathrm{S}j}\hat{\sigma}^{+}_{j}\hat{E}_{\mathrm{S}j}(\text{y})e^{-i\Delta t} \bigg]\nonumber\\
&&+ H.c.
\eea
In writing Eq. (\ref{eq5}) we have for simplicity considered the transition frequencies of the emitters to be identical such that  $\Delta=(\omega_{0}-\omega_{eg})$ and have dropped the subscript $k$ from $\mathcal{G}$ by assuming that the coupling strengths are similar for the continuum modes. 

We consider the initial state of the emitter-field system to be a coherent state of the form  
\be
\label{eq6}
\Ket{\Psi(t=-\infty)}=\hat{D}(\alpha_{k})\prod_{j}\Ket{g_{j}}\Ket{\emptyset},
\ee
where the emitters are in the ground state $|g_{j}\rangle$ while the input is in a coherent state of amplitude $\alpha_{k}$ represented by the standard interaction picture coherent state displacement operator $\hat{D}(\alpha_{k})=\mathrm{e}^{\left(\int d k\left(\hat{a}_{\text{R},k}^{\dagger}\alpha_{k}-\alpha_{k}^{*}\hat{a}_{\text{R},k}\right)\right)}$ acting on the vacuum state $\Ket{\emptyset}$. Note that in defining the displacement operator we have explicitly assumed that the input field is incident from the left end of the waveguide and is propagating towards the right. 

We next perform a frame transformation with respect to the displacement operator such that $\Ket{\tilde{\Psi}(t)}=\hat{D}^{\dagger}(\alpha)\Ket{\Psi(t)}$. On using the standard property of the displacement operator \cite{Scullyb}
\bea
\label{a2}
\hat{D}^{\dagger}(\alpha_{k'})\int\mathrm{d}k\hat{E}_{k}^{\dagger}\hat{D}(\alpha_{k'})=\int\mathrm{d}k\left(\hat{E}^{\dagger}_{k}+\mathcal{E}_{k}^{*}\right),
\eea
the transformed interaction Hamiltonian becomes $\hat{\tilde{\mathcal{H}}} = \sum_{j}\hat{\tilde{\mathcal{H}}}_{j} = \hat{D}^{\dagger}(\alpha_{k})\sum_{j}\hat{\mathcal{H}}_{j}\hat{D}(\alpha_{k})$ \cite{ Mollow}, where 
\bea
\label{eq7}
\hat{\tilde{\mathcal{H}}}&=&-\hbar\sqrt{2\mathrm{\pi}}\bigg[\sum_{j = 1}^{N}\int d\text{z}~\delta(\text{z}-\text{z}_{j})~\hat{\sigma}^{+}_{j}\bigg(\mathcal{G}_{\mathrm{R}j}\hat{E}_{\mathrm{R}}(\text{z})e^{-i(\Delta t-k_{0}\text{z})}\nonumber\\
&+&\mathcal{G}_{\mathrm{L}j}\hat{E}_{\mathrm{L}}(\text{z})~e^{-i(\Delta t+k_{0}\text{z})}+\mathcal{G}_{\mathrm{R}j}\mathcal{{E}}e^{-i(\Delta t-k_{0}\text{z})}\bigg)+\sum_{j = 1}^{N}\int~dy\nonumber\\
&\times&\delta(y)\mathcal{G}_{\mathrm{S}}\sigma^{\dagger}_{j}\hat{E}_{\text{S}j}(\text{y})e^{-i\Delta t}\bigg] + H.c.,
\eea
where $\mathcal{{E}}=\int\mathrm{d}k\alpha_{k}e^{i[(k-k_{0})\text{z}+(\omega_{0}-\omega_{k})t]}$ corresponds to the coherent input field amplitude at position $\text{z}$ and time $t$. The above transformation is essentially a change of basis that maps the initial state of the system to vacuum, while the right-propagating photon operator transforms as $\hat{D}^{\dagger}(\alpha_{k})\hat{E}_{\mathrm{R}}(\text{z})\hat{D}(\alpha_{k})\rightarrow\hat{E}_{\mathrm{R}}(\text{z})+\mathcal{{E}}$. The advantage of this transformation is that we do not explicitly need to include the initial quantum states of the field in the dynamics. This substantially simplifies the wave-function ansatz considered in the following section as now the initial state of the field is simply represented as a (classical) driving field. Note further, that we here for concreteness consider the field to be in the waveguide. In experiments the systems is often driven by external fields not inside the waveguide. In this case the resulting Hamiltonian above is exactly the same, and one should below just ignore the contribution of the coherent state for calculating the field in the forward direction. 
\begin{figure}[h!]
	\includegraphics[height = 4 cm]{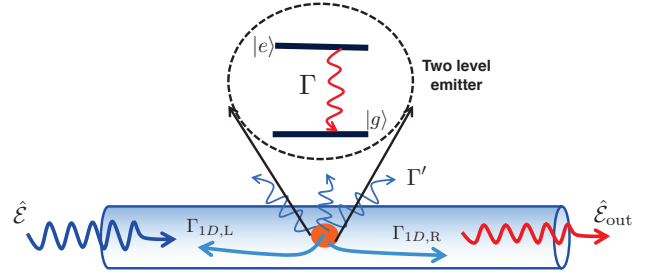}
	\caption{Schematic of a single  two-level quantum emitter coupled to the field mode of a $1$D waveguide and interacting with the incoming field $\hat{\mathcal{E}}$ with the emitted field from the quantum emitter is represented by $\hat{\mathcal{E}}_{\text{out}}$ respectively. \label{figb}}
\end{figure}
\section{Correlations of photons emitted by a single two-level emitter}

Now that we have described the general form of the Hamiltonian governing the systems we develop our formalism in this section by first investigating the correlation properties of photons emitted from the simple system of a single two-level emitter coupled to a $1$D waveguide shown schematically in Fig.(\ref{figb}). In section V. we will generalize the formalism developed here to make it applicable for study of quantum dynamics of multi-emitter systems.
\subsection{The Wave function Ansatz}
To study the dynamics of scattering we first introduce an emitter-field wave function ansatz. The wave function ansatz contains all possible states with up to two excitations and has the form \cite{Chang07, Mantas_thesis, Liang_thesis}
\begin{widetext}
	\bea
	\label{eq25}
	\Ket{\tilde{\Psi}(t)} &=&c_{g}(t)\Ket{g, \emptyset}+c_{e}(t)\Ket{e, \emptyset}+ \int dt_{e}~\phi_{g\text{R}}(t,t_e)\hat{E}_{\text{R}}^{\dagger}(v_g(t-t_e))\Ket{g,\emptyset}+\int dt_e~\phi_{e\text{R}}(t,t_e)\hat{E}_{\text{R}}^{\dagger}(v_g(t-t_e))\Ket{e,\emptyset}\nonumber\\
	&+&\int dt_{e}~\phi_{g\text{S}}(t,t_e)\hat{E}_{\text{S}}^{\dagger}(v_g(t-t_e))\Ket{g,\emptyset}+\int dt_{e}~\phi_{e\text{S}}(t,t_e)\hat{E}_{\text{S}}^{\dagger}(v_g(t-t_e))\Ket{e,\emptyset}+\int dt_{e2}\int dt_{e1}~\phi_{\text{RR}}(t,t_{e2},t_{e1})\nonumber\\
	&\times&\hat{E}_{\text{R}}^{\dagger}(v_g(t-t_{e2}))\hat{E}_{\text{R}}^{\dagger}(v_{g}(t-t_{e1}))\Ket{g,\emptyset}+\int dt_{e2}\int dt_{e1}~\phi_{\text{RL}}(t,t_{e2},t_{e1})\hat{E}_{\text{R}}^{\dagger}(v_g(t-t_{e2}))\hat{E}_{\text{L}}^{\dagger}(v_g(t-t_{e1}))\Ket{g, \emptyset}\nonumber\\
	&+&\int dt_{e2}\int dt_{e1}~\phi_{\text{RS}}(t,t_{e2},t_{e1})\hat{E}_{\text{R}}^{\dagger}(v_g(t-t_{e2}))\hat{E}_{\text{S}}^{\dagger}(v_g(t-t_{e1}))\Ket{g, \emptyset}+
	\int dt_{e2}\int dt_{e1}~\phi_{\text{SR}}(t,t_{e2},t_{e1})\hat{E}_{\text{S}}^{\dagger}(v_g(t-t_{e2}))\nonumber\\
	&\times&\hat{E}_{\text{R}}^{\dagger}(v_g(t-t_{e1}))\Ket{g, \emptyset}
	+\int dt_{e2}\int dt_{e1}~\phi_{\text{SS}}(t,t_{e2},t_{e1})\hat{E}_{\text{S}}^{\dagger}(v_g(t-t_{e2}))\hat{E}_{\text{S}}^{\dagger}(v_{g}(t-t_{e1}))\Ket{g,\emptyset}+\text{R}\leftrightarrow\text{L}.
	\eea
\end{widetext}
Here $c_{g}$ and $c_{e}$ are respectively the amplitude of the ground state $|g\rangle$ and excited state $|e\rangle$ with the field state being vacuum $|\emptyset\rangle$. The terms $\phi_{g (e) \text{R(L)}}(t,t_e)$ represent the  amplitude of the combined emitter-field state, with the emitter being in the ground (excited) state at a time $t$ with a right (left)-propagating photon emitted at time $t_{e}$, while the terms $\phi_{g (e)\text{S}}(t,t_e)$ corresponds to the amplitude of the emitter-field state when a photon is lost out of the waveguide at some time $t_s$.  Furthermore, the amplitude $\phi_{\text{RR(LL)}}(t,t_{e2},t_{e1})$ corresponds to the field state with all the emitters in their ground state, when two right (left)-propagating photons are emitted at times $t_{e1}$ and $t_{e2}$ by the emitter, while $\phi_{\text{SS}}(t,t_{e2},t_{e1})$ is the amplitude when both the photons are lost to the outside of the waveguide at time $t_{e2}$ and $t_{e1}$. Finally, $\phi_{\text{RL(LR)}}(t,t_{e2},t_{e1})$ represents the amplitude of the field state with the emitter being in the ground state and photons propagating along opposite directions. Here we assume the first photon being emitted at time $t_{e1}$ propagating right (left) while the second one emitted at $t_{e2}$ propagating towards left (right), respectively. In addition, the terms $\phi_{\text{R(L)S}}(t,t_{e2},t_{e1})$ and $\phi_{\text{SR(L)}}(t,t_{e2},t_{e1})$ correspond to the state amplitude when one emitted photon at time $t_{e1}$  is propagating to the right (left), while the other is lost out of the waveguide at $t_{e2}$ and vice versa. Note that in this work for all the two photon processes we have explicitly considered $t_{e2} > t_{e1}$.

In writing the above wave-function ansatz we consider a single mode for the output, for which the spatially dependent field operators can be defined as done in Eq. (\ref{eq4}) thereby allowing the photons to be readily tracked by their position in the waveguide. Furthermore, to keep the description tractable, we truncate the Hilbert space and exclude states with three or more excitations. This truncation is physically motivated by considering systems where the probability for emitting more than two excitations is almost negligible. This is, e.g., the case if the incident field is suffciently weak or if the system does not have time to emit more than two excitations. The approach can easily be extended to include more excitations  but in this case the complexity grows and it may be desirable to resort to other means, e.g., Monte Carlo wavefunctions. 

We have written the co-efficients of the states in the wave-function in Eq. (\ref{eq25}) with explicit dependence on both the emission time $t_{e}$ and the time $t$. This is to describe that there are various times for the dynamics of the system, like,  emission of photons and dynamics of the system both, occurring after the emission. If $t<t_{e}$, no photon has yet been emitted and thus the amplitude vanishes $\phi_{e\text{R}}(t,t_e) = 0$. After the emission at time $t_e$  the emitter may still evolve, e.g., get excited from $|g\rangle$ to $|e\rangle$, resulting in an additional time dependence $t$. Since the photon may be emitted at all possible times the total state is a superposition over all possible emission times as described by the time integral over the emission(s) at times $t_e, t_{e1}$ and $t_{e2}$.  

The wavefunction ansatz in Eq. (\ref{eq25}) is motivated by physical considerations, i.e. it contains amplitudes corresponding to the process one would expect in the dynamics. As we will now show, this ansatz does indeed provide an accurate description of the systems provided the number of excitations in the system is less than or equal to two.

\subsection{The field-emitter dynamics}
We next derive the equation of motion for the probability amplitudes by invoking the time dependent Schr\"{o}dinger equation $\partial\tilde{\Psi}/\partial t = (-i/\hbar)\tilde{\mathcal{H}}\tilde{\Psi}$. The details of this derivation is presented in appendix A. Here we only discuss the important assumptions and considerations that lead us to the correct description. We make a key observation about the equations of motion for $\phi_{g\text{R}}$ and $\phi_{g\text{L}}$ in appendix A. Both of them vanish at the initial time $t = -\infty$ since no photon has yet been emitted and they are part of a homogeneous set of differential equations apart from a source term $c_{e}(t_e)\delta(t-t_e)$. This means that these two state amplitudes get a contribution from the single excitation amplitude only after the emission happens i.e. after the evolution time $t$ has crossed the emission time $t_{\mathrm{e}}$  and vanish at earlier times. This observation has important consequence for the state amplitudes. For example, all the other state amplitudes, which describe states with emitted photons, vanish before the emission and only acquire amplitudes through processes such as $\phi_{g\text{R(L)}}\rightarrow\phi_{e\text{R(L)}}\rightarrow\phi_{\text{R(L)R(L)}}$. We use this fact to split the dynamics of the system into two different time windows, $0< t< t_{e}-\varepsilon$ and $t_e+\varepsilon< t <\infty$, where $\varepsilon$ is an infinitesimal time period. 

As shown in appendix A, the coupled set of differential equations involving all the probability amplitudes then decouple into three sets of coupled differential equations. The first set involves amplitudes $c_{g}$, $c_{e}$, while the second set involves $\phi_{g\text{R(L)}}$ and $\phi_{e\text{R(L)}}$ and the last set involves $\phi_{\text{R(L,S)R(L,S)}}$, which does not evolve after the emission since we truncate the dynamics to two excitations, and this state already contains two emitted photons. This simplifies the analytical treatment of the dynamics substantially. From the Schr\"{o}dinger equation we get two sets of equations. One concerning the dynamics before emission, $(0 <\zeta< \zeta_{e})$,
\bea
\label{eq31}
\dot{c}_{g}(\zeta) & = &i\frac{\Omega}{2}^{\ast}e^{i\tilde{\Delta}\zeta}c_e(\zeta),\\
\label{eq32}
\dot{c}_{e}(\zeta) & = &i\frac{\Omega}{2}e^{-i\tilde{\Delta}\zeta}c_{g}(\zeta)-\frac{1}{2}c_e(\zeta).
\eea
while the other corresponds to the dynamics after emission of a photon $(\zeta_{e} <\zeta< \infty)$ and is given by,
\bea
\label{eq38}
\dot{\tilde{\phi}}_{gj}(\zeta,\zeta_{e})& = &i\frac{\Omega}{2}^{\ast}e^{i\tilde{\Delta}\zeta}\tilde{\phi}_{ej}(\zeta,\zeta_{e}),\\
\label{eq39}
\dot{\tilde{\phi}}_{ej}(\zeta,\zeta_{e})& = &i\frac{\Omega}{2}e^{-i\tilde{\Delta}\zeta}\tilde{\phi}_{gj}(\zeta,\zeta_{e})-\frac{1}{2}\tilde{\phi}_{ej}(\zeta,\zeta_{e}),
\eea
with the suffix $j = \{L,R, S\}$. In writing these expressions, we have made all the physical variables dimensionless by normalizing with respect to the total decay rate $\Gamma = \Gamma_{1D,\text{R}}+\Gamma_{1D,\text{L}}+\Gamma'$, of the excited state of the two-level emitter. Here $\Gamma_{1D,\text{R(L)}} = 2\pi \mathcal{G}_{\text{R(L)}}^{2}/v_g$ and $\Gamma' = 2\pi\mathcal{G}_{\text{S}}^{2}/{c}$ are respectively the decay rate into the waveguide in the right (left) propagating mode and to the outside. The time $t$ is then replaced by a new variable $\zeta$ defined by $\zeta=\Gamma t$ (the time of emission similarly becomes dimensionless following the definition $\zeta_e = \Gamma t_e$). The dimensionless Rabi frequency is defined by $\Omega = \sqrt{\beta_{R}}\tilde{\mathcal{E}}e^{ik_{0}\text{z}}$, where $\mathcal{{\tilde{E}}} = \Gamma \mathcal{{E}}/v_{g}$, $\tilde{\phi}_{e/g\text(R/L)} = \phi_{e/g\text(R/L)}/\sqrt{\Gamma v_{g}}$, $\tilde{\phi}_{\text{R(L)R(L)}}= \phi_{\text{R(L)R(L)}}/(\Gamma v_{g})$ and the detuning is renomalized to $\tilde{\Delta} = \Delta/\Gamma$. All other physical variables are also normalized with respect to $\Gamma$, giving new dimensionless variables. We have also used the relation $\beta = \beta_{R}+\beta_{L}+\beta_{s} = 1$ to get the factor $1/2$ in front of $c_e(\zeta)$ in Eq. (\ref{eq32}) with the definition $\beta_{j} = \Gamma_{j}/\Gamma$ i.e. $\beta_j$ is the branching ratio for an emitter to decay through the channel $j$.

From Eqs. (\ref{eq31}) and (\ref{eq32}) we find that there are only two processes that affect the dynamical evolution of emitter states until the emission of a photon: $(1)$ excitation of the ground state by the input field (the terms proportional to $\mathcal{{E}}$ in Eq. (\ref{eq31})), and $(2)$ the decay of the excited state into any channel (given by $-\frac{1}{2}c_e$). Furthermore, the equation of motion (\ref{eq38}) and (\ref{eq39}) for the field amplitudes after decay are identical to the equations representing the dynamics of the emitter states before the decay (Eq. \ref{eq31} and \ref{eq32}). This is expected since the equations are derived within the Markov approximation where the environment has no memory of the previous evolution. Once a photon has been emitted the system thus evolves in the same manner as before the emission. 

Eqs. (\ref{eq31}-\ref{eq39}) conveniently describe the evolution before or after the emission. To fully describe the dynamics we, however, also need the initial condition. This is given by noting that $\phi_{e/gj}(\zeta,\zeta_{e})=0$ for $\zeta <\zeta_{e}$ along with the condition given by, 
\bea
\label{eq30}
\tilde{\phi}_{g\text{j}}(\zeta_{e}+\varepsilon,\zeta_e)  =i\sqrt{\beta_{j}}c_e(\zeta_{e})e^{i\tilde{\Delta}\zeta_{e}}.
\eea
describing that the amplitude for single photon emission is proportional to the amplitude of being in the excited state before the emission. Similarly for the two photon emission we have the initial condition that it vanish unless $\zeta > \zeta_{e2} > \zeta_{e1}$ along with the condition 
\bea
\label{eq30a}
\tilde{\phi}_{jk}(\zeta_{e_2}+\varepsilon,\zeta_{e_2},\zeta_{e_1})  = i \sqrt{\beta_{j}}\tilde{\phi}_{ek}(\zeta_{e2}, \zeta_{e_1})e^{i\tilde{\Delta}\zeta_{e2}}.
\eea
which state that the amplitude of two photon emission is proportional to the amplitude of first having emitted a photon and then having the emitter excited. Note that these two photon components have been integrated and their solution substituted in the equations of motion for $\tilde{\phi}_{\text{eR}}$ (see appendix A), to get the present form of Eq. (\ref{eq39}). 

\subsection{Correlation properties of scattered photons}
In this section, we investigate how to evaluate the correlations between the emitted photons. For this purpose, we consider the normalized second-order correlation function $g^{(2)}(\tau,t)$ which is defined as \cite{Scullyb, Ficekb} 
\be
\label{eq18}
g^{(2)}(\tau, t) = \frac{\text{G}^{(2)}(t+\tau,t)}{\text{G}^{(1)}(t+\tau)\text{G}^{(1)}(t)}. 
\ee
Here we have redefined the first order correlation function $\text{G}^{(1)}(t, t) = \text{G}^{(1)}(t) = \langle \text{I}(t)\rangle$ in terms of the average intensity and used the standard definition of the second order correlation function $\text{G}^{(2)}(t+\tau,t) = \langle: \text{I}(t+\tau)\text{I}(t):\rangle$ as the normal ordered correlation between the intensity of photons emitted at time $t$ and $t+\tau$. There are certain notable properties of the $g^{(2)}$ function. For example, $g^{(2)}(0) = g^{(2)}(\tau) = 1$, correspond to uncorrelated light while $g^{(2)}(0) > 1$ signifies photon bunching effects. For problems concerning emission and detection of single photons, the most notable characteristic is the condition $g^{(2)}(0) < 1$, with $g^{(2)} (0) = 0$ being the signature of ideal single photon emission.   

To evaluate the first- and second-order correlations we face the formal problem that an operator describing the time of arrival of a photon cannot be defined in quantum mechanics. We, however, consider photons propagating with constant group velocity and we can therefore avoid this problem by instead assuming that at some time after the photons have left the system, we measure the position of the photons. We are interested in evaluating the intensity correlation $\langle :\text{I}(t_d+\tau_d)\text{I}(t_d):\rangle$ at some detection time $t_{d}$ for a detector placed at a position $\text{z}_{D}$ to the right of the emitter. The temporal correlation function can then be evaluated by a spatial correlation function
\begin{widetext}
\bea
\label{eqc11}
\langle: \text{I}(t_d+\tau_d)\text{I}(t_d):\rangle =v_g^2\langle E_\text{R}^\dagger(\text{z}_D+v_g(T-t_d- \tau_d ))E_\text{R}^\dagger(\text{z}_D+v_g(T-t_d))E_\text{R}(\text{z}_D+v_g(T-t_d))E_\text{R}(\text{z}_D+v_g(T-t_d-\tau_d ))\rangle,\nonumber\\
\eea
\end{widetext}
at some later time $T$. Here we have complete freedom in choosing $T$ as long as $T>t_d, t_d+\tau_d$ to ensure that the right going field is evaluated to the right of the ensemble. This freedom translates into a freedom in how the final expressions are evaluated. Note that in writing Eq. (\ref{eqc11}), for brevity we have not used the transformation of the field operators given in Eq. (\ref{a2}). 

Using the expression in Eq. (\ref{eqc11}), the photon correlations can be evaluated in terms of the state amplitudes of the wave-function given in Eq. (\ref{eq25}). In presence of an incoming right propoagating field $\mathcal{E}$, we need to perform the transformation described in  Eq. (\ref{a2}) so that Eq. (\ref{eqc11}) and the corresponding first order correlation functions can be written as
\bea
\label{eq19}
&&\text{G}^{(1)}(t_d) = v_{g}\langle\tilde{\Psi}(T)|\left(\hat{E}^{\dagger}_{\text{R}}(\text{z}_{t_d})+\mathcal{E}^\ast\right)\left(\hat{E}_{\text{R}}(\text{z}_{t_d})+\mathcal{E}\right)\Ket{\tilde{\Psi}(T)},\nonumber\\
\eea
\begin{widetext}
\bea
\label{eq20}
&&\text{G}^{(2)}(t_d+\tau_d,t_d) = v^{2}_{g}\Bra{\tilde{\Psi}(T)}\left(\hat{E}^\dagger_{\text{R}}(\text{z}_{t_d+\tau_d})+\mathcal{E}^\ast\right)\left(\hat{E}^\dagger_{\text{R}}(\text{z}_{t_d})+\mathcal{E}^\ast\right) \left(\hat{E}_{\text{R}}(\text{z}_{t_d+\tau_d})+\mathcal{{E}}\right)\left(\hat{E}_{\text{R}}(\text{z}_{t_d})+\mathcal{{E}}\right)\Ket{\tilde{\Psi}(T)}
\eea
\end{widetext}
where $\text{z}_{t_d} = v_{g}\left(T-t_d\right)+\text{z}_{D}$. Note that the wave-function in the above expression is evaluated at a time $T$ such that the relevant photon wave-packet is to the right of the ensemble.

The first and second-order correlation functions of Eq. (\ref{eq19}) and Eq. (\ref{eq20}) for photons  scattered from a single two level emitter is then evaluated in dimensionless variables to be 
\bea
\label{eq40}
\text{G}^{(2)}(\zeta'_e,\zeta_e) & =& v_{g}^{2}\bigg|\tilde{\phi}_{\text{RR}}(\zeta_{T},\zeta'_{e},\zeta_{e})+\tilde{\phi}_{\text{RR}}(\zeta_{T},\zeta_{e},\zeta'_{e})\nonumber\\
&&+\mathcal{\tilde{E}}(\zeta'_{e})\tilde{\phi}_{g\text{R}}(\zeta_{T},\zeta_{e})+\tilde{\phi}_{g\mathrm{R}}(\zeta_{T},\zeta'_{e})\mathcal{\tilde{E}}(\zeta_{e})\nonumber\\
&&+\mathcal{\tilde{E}}(\zeta'_e)\mathcal{\tilde{E}}(\zeta_e)\bigg|^{2},\\
\label{eqc2}
\text{G}^{(1)}(\zeta_e) &  = &v_{g}\left|\tilde{\phi}_{g\text{R}}(\zeta_{T},\zeta_{e})+\mathcal{{\tilde{E}}}(\zeta_e)\right|^{2},
\eea
where $\zeta_{e}$ is the dimensionless variable representing the emission time, which is related to the detection time by $\zeta_{e} = \left(\zeta_{d}-\frac{\Gamma \text{z}_{D}}{v_{g}}\right)$. Here the term $\Gamma \text{z}_{D}/v_{g}$ merely reflects the retardation in going from the emitter to the detector, which just offsets the measured time due to the propagation of the light. Hence it is more convenient to express the correlation functions in terms of the emission time $\zeta_e$ rather than the detection time $\zeta_d$.  The other dimensionless variable $\zeta_{T}$ corresponds to some time $T$ after the emission. The photon correlation functions above has been derived in a weak field consideration ($\mathcal{E} << 1$) and hence have only the leading order terms. For the full expression of the correlation functions in all orders of $\mathcal{E}$ please refer to appendix C. 
            
\begin{figure*}
	\centering{}\includegraphics[scale=0.4]{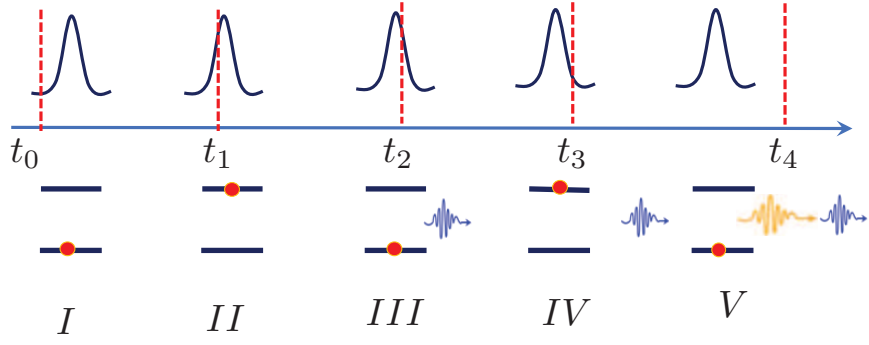}\caption{Schematics of two-photon emission from a two-level quantum system that spoils the quality of a single photon source.  I: Before $t_0$, no driving laser is applied and the two level quantum emitter is in the ground state. II: within the pulse duration at $t_1$  the emitter is transferred to the excited state.  III: The emitter decays via spontaneous emission at $t_2$ and a narrow single-photon (blue) is emitted before the tail of the excitation pulse have passed . IV-V: The pulsed laser being still present interacts with the emitter further and excites it again at $t_3$, which then leads to a second photon emission via spontaneous decay (yellow) after the pulse duration. \label{fig5}}
\end{figure*}
                                                                                                                                                             
\section{Application of the wave-function ansatz method to the study of spectral filtering of photons}
In this section we apply the above developed method, to study effect of spectral filtering on correlations of  
photons emitted from a quantum emitter coupled to a $1$D waveguide. To begin with, we discuss how two photon emission can occur within an excitation pulse length and how the corresponding $g^{(2)}$ for a pulsed excitation can be evaluated using our method.
\subsection{Second order correlation function for pulsed excitation of a quantum emitter }
An ideal single-photon source is expected to emit a single-photon in a deterministic manner at periodic intervals. This can in principle be achieved by pulsed resonant excitation of a two level quantum emitter. Ideally, a pulsed laser excites the emitter arbitrarily fast to the excited state and then a single-photon is emitted via spontaneous emission. In practice, however, due to finite excitation pulse width and limited excitation power the quantum emitter can be excited twice and emit two photons within the pulse duration as shown schematically in Fig \ref{fig5}. In this case, the emitter decays to the ground state within the pulse duration and since the excitation process is still active, the emitter can then get excited a second time and radiatively decay back to the ground state (see Fig \ref{fig5} (IV-V)). This leads to a finite probability for two-photon emission within one pulse interval which in turn makes single-photon generation impure and renders the source imperfect \cite{Lukas17, Liang_thesis, Fisher16}. We would like to re-emphasize that there can also be other processes like three or more photons emitted under one excitation pulse, but the possibilities of those are much lower and we can neglect them for the present discussion. This makes the system highly suited for our method, which neglects such higher order states. 

The probability of double excitation is related to the properties of the excitation pulse and the radiative decay rate. For a certain pulse shape and a certain excitation power, the ratio between the excitation pulse width and the radiative lifetime of the emitter determines the probability of two photon emission within the pulse duration. In analogy to the continuous time correlation function in Eq. (\ref{eq18}), the purity of a single photon pulse is characterized by the second order correlation function given by \cite{Liang_thesis, Kiraz04}
\bea
\label{eq46}
g^{(2)}_{p} & = & \frac{\text{G}^{(2)}_{\text{pulsed}}}{\langle n\rangle^{2}}\nonumber\\
& = &\frac{\int^{T_\text{p}}_{0}d\tau \int^{T_\text{p}}_{0}dt\langle \hat{E}^{\dagger}(t)\hat{E}^{\dagger}(t+\tau)\hat{E}(t+\tau)\hat{E}(t)\rangle}{(\int^{T_\text{p}}_{0}dt\langle \hat{E}^{\dagger}(t)\hat{E}(t)\rangle)^{2}}\nonumber\\
\eea
Here $T_{\text{p}}$ represents the duration of the excitation pulse and $\langle n\rangle$ is the mean number of photons in the pulse. For pure single photon emission, we expect a vanishing second-order correlation function, i.e. $g^{(2)}_{\text{p}} = 0$. In practise however, because of the above discussed two photon emission events, $g^{(2)}_{\text{p}}$ attains a finite value. Since the amplitude of the double excitation process is related to the excitation pulse width and the excitation power, the value of $g^{(2)}_{\text{p}}$ is also expected to depend on them. To minimize the probability of two-photon emission, a small ratio between the excitation pulse width and the emitters lifetime is preferred. However, finite pulse duration and laser power requirements will always introduce some imperfection. 

To evaluate the performance of a single photon source we consider a single two level quantum emitter coupled to a $1$D waveguide and driven by a laser incident from the side such that no residual excitation light hits the detector. We can thus ignore the incident field $\mathcal{E}$ for calculating the outgoing field. The dynamics of the emitter is then given by the set of equations (\ref{eq31}-\ref{eq39}) derived in section III. B. The total number of photons detected during a pulse duration $T_{p}$ is then given as $\langle n\rangle = \int^{v_{g}T}_{v_{g}(T-T_{p})} d\text{z}~\langle\hat{E}^\dagger(\text{z})\hat{E}(\text{z})\rangle$, where we have assumed that the wave function is evaluated at a final time $T$.

\begin{figure}
	\centering{}\includegraphics[scale=0.6]{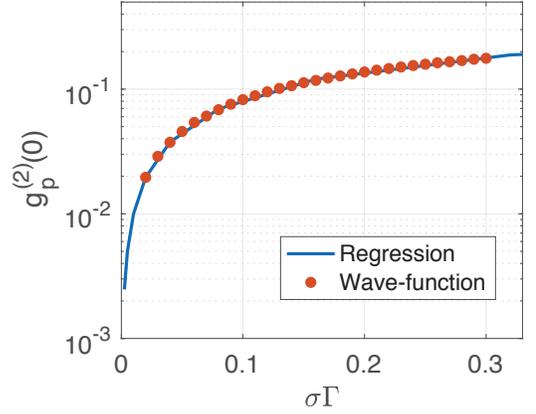}\caption{Pulse-width dependence of the second order correlation function $g^{(2)}_{p}$ simulated with quantum regression method and the introduced wave-function ansatz method, plotted on a logarithmic scale. The pulse width axis is normalised by the lifetime of the emitter. The solid line (blue) corresponds to the $g^{(2)}_{p}$ simulated using quantum regression theory while the circles (red) are obtained using the wave-function method. \label{fig6}}
\end{figure}

Following the analytical treatement of scattering from such single emitter system detailed in Sec. III, and on using Eq. (\ref{eq40}) we find the pulsed second-order correlation function $G^{(2)}_{\text{pulsed}}$ in dimensionless co-ordinates to be
\bea
\label{eq48}
\text{G}^{(2)}_{\text{pulsed}} = 2v^{2}_{g} \int^{\zeta_T}_{0}~d\zeta'_{e}\int^{\zeta'_{e}}_{0}~d\zeta_{e}~|\tilde{\phi}_{\text{RR}}(\zeta_{T},\zeta'_{e},\zeta_{e})|^{2}. \nonumber\\
\eea
Similarly, we evaluate $\langle n \rangle$ in terms of first order correlation function given in Eq. (\ref{eqc2}) as
\bea
\label{eq49}
\langle n\rangle & = &v_{g}\bigg\{\int^{\zeta_{T}}_{0}d\zeta'_e|\tilde{\phi}_{g\text{R}}(\zeta,\zeta_{e})|^{2}+2\int^{\zeta_T}_{0}d\zeta'_{e}\int^{\zeta'_{e}}_{0}d\zeta_{e}\nonumber\\
&\times&|\tilde{\phi}_{\text{RR}}(\zeta_{T},\zeta'_{e},\zeta_{e})|^{2}\bigg\}.
\eea                
Note that, for evaluation of the correlation functions of emitted photons from a single emitter, the distinction between $\beta_\text{R}$, $\beta_\text{L}$ and $\beta_\text{S}$ is not important, since this just gives a constant branching rate between the various decay paths. Since $g^{(2)}_{p}$ is insensitive to such branching ratios, we for simplicity evaluate it using $\beta_\text{R} = 1$ and neglect the left and side channel completely.  The normalised second-order correlation function defined in Eq. (\ref{eq46}) can be obtained using Eqs. (\ref{eq48}) and (\ref{eq49}). We plot $g^{(2)}_{p}(0)$ in a semi-logarithmic scale in Fig. \ref{fig6} as a function of normalized pulse width. The simulation is performed with an input Gaussian pulse of the form $\Omega^{2}(t) = A/\sqrt{2\pi\sigma} \exp[-(t)^{2}/2\sigma^{2}]$, where $A$ is the normalization factor such that $\int \Omega(t)dt = \pi$ and $\sigma$ is the width of the pulse with a full width at half maximum (FWHM) of $\sim 2.35 ~\sigma$.

To check the validity of our wave-function approach, we compare our evaluated second order correlation function with that obtained from the standard quantum regression method \cite{Ficekb}. In Fig. \ref{fig6}, the solid line (blue) corresponds to $g^{(2)}_{p}$ calculated via the quantum regression method, while the circles (red) are evaluated using the wave-function method introduced in this work. The figure clearly shows that our introduced method matches exactly to that of the quantum regression theorem. For a single emitter our method has no particular advantage over the quantum regression theorem in the calculation of $g^{(2)}$. For the filtering considered below, however, the wave function method provides a major advantage. 

\subsection{Spectral filtering of photons} 
To theoretically study the effect of spectral filtering, the spectrum for both single-photon emission and two-photon emission has to be calculated. The ability to separate the single-photon and two-photon components using the introduced time-domain wave-function method allows us to calculate the effect of spectral filtering on the second-order correlation function and thereby on the purity of the emitted single photons. As pointed out already, for a general filter, solutions of this problem via the quantum regression theorem are not tractable owing to the need to calculate four point correlation functions. 
\begin{figure*}
	\centering{}\includegraphics[scale=0.5]{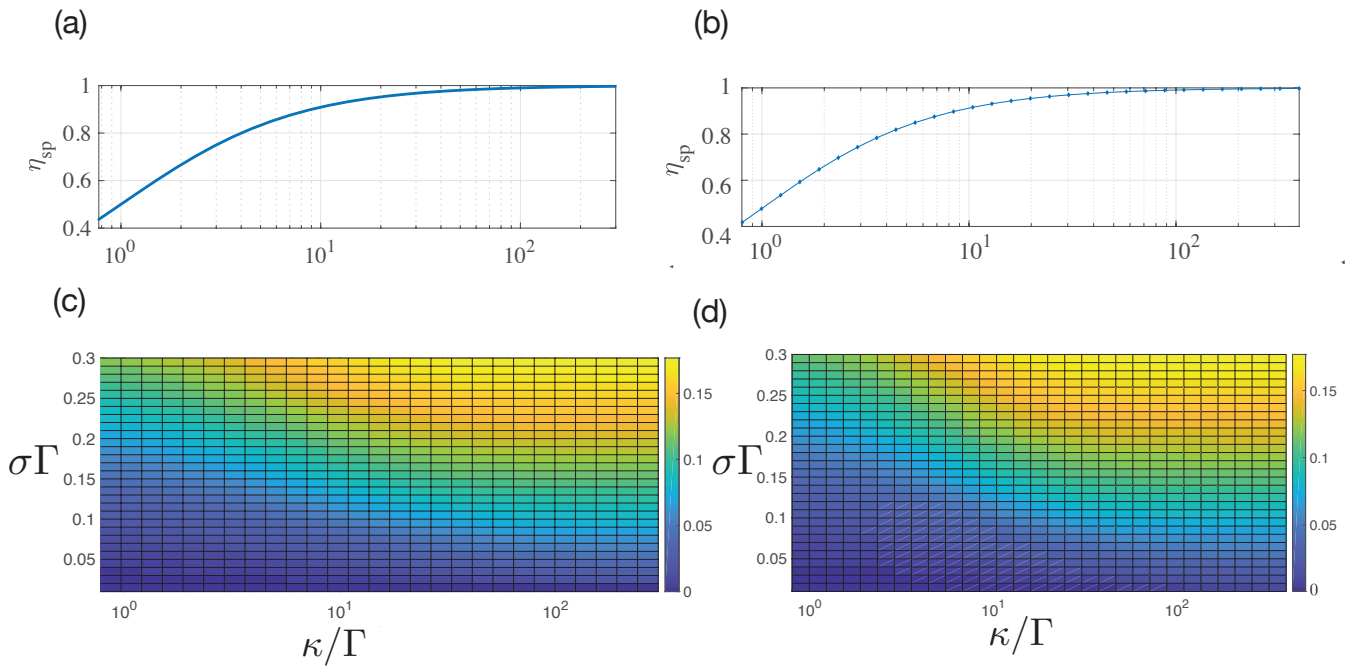}
	\caption{Single-photon transmission efficiency $\eta_{\text{sp}}$ in the presence of (a) a Lorentzian filter and (b) a Gaussian filter, respectively.  The $2$D colour coded map shows the effect of filters on $g^{(2)}$ for a (c) Lorentzian and (c) Gaussian filter, respectively. \label{fig8}}
\end{figure*}

To investigate the filtering process we first introduce the relation between frequency domain input and output field mode operators in the Heisenberg picture \cite{Gard85}
\bea
\label{eq53}
\hat{E}_{\text{out}}(\omega) = \mathcal{T}(\omega-\omega_{\text{c}})\hat{E}_{\text{in}}(\omega),
\eea
where $\mathcal{T}(\omega - \omega_{\text{c}})$ is the amplitude transmission function of the filter with $\omega_{c}$ being the cavity frequency.

The above derived input-output relation for the filtering process involves the photon mode operators and is derived in the Heisenberg picture, while our wave-function method is based on the Schr\"{o}dinger picture. To combine these two approaches, and analyze the filtering process let us first derive a relation between the description of photon mode operators in space and time domain. A photon arriving at the filter at a time $t$ can be written as 
\bea
\label{eq55}
\hat{E}(\text{z} = \text{L}_{1}, t) & = & \hat{E}_{\text{in}}(\text{z} = \text{L}_{1}-v_{g}(t-T)),
\eea
where we have assumed that the entry point of the filter have a spatial position $\text{z} = \text{L}_{1}$ with $T$ being some initial time for the photon to the left of the filter $(\text{z} < \text{L}_{1})$ and $\hat{E}_{\text{in}}$ is the field at that time defined as $E_{\text{in}}(\text{z}) =  E(\text{z},T)$. On using the frequency domain definition of the $\text{G}^{(2)}_{\text{pulsed}} =v^{2}_{g} \int~d\omega_{1}~\int~d\omega_{2}~ \langle \tilde{\Psi}| ~\hat{E}^{\dagger}_{\text{out}}(\omega_{1})\hat{E}^{\dagger}_{\text{out}}(\omega_{2})\hat{E}_{\text{out}}(\omega_{2})\hat{E}_{\text{out}}(\omega_{1})~|\tilde{\Psi}\rangle$ corresponding to Eq. (\ref{eq48}), and substituting the cavity input-output relation from Eq. (\ref{eq53}) we get the unnormalized second order correlation function in terms of the filter function and the two-photon amplitude as
\bea
\label{eq56}
\text{G}^{(2)}_{\text{pulsed}}& = &\frac{v^{2}_{g}}{4\pi^2}\int^{\infty}_{-\infty} d\omega_{1}\int^{\infty}_{-\infty} d\omega_{2} ~|\mathcal{T}({\omega_1})|^{2}|\mathcal{T}(\omega_{2})|^{2}\nonumber\\
&\times&\bigg|\int^{\zeta_T}_{0} d\zeta'_{e}\int^{\zeta'_e}_{0} d\zeta_{e}~\tilde{\phi}_{\text{RR}}(\zeta_{T},\zeta'_{e}, \zeta_{e})\nonumber\\
&\times&\left(e^{-i(\omega_{1}\zeta'_{e}+\omega_{2}\zeta_{e})}+e^{-i(\omega_{1}\zeta_{e}+\omega_{2}\zeta'_{e})} \right)\bigg|^{2},
\eea
where we have used the Fourier transformation of $\hat{E}_{\text{in}}(\omega)$ in deriving the above expression. The integration inside the absolute value sign in Eq. (\ref{eq56}) can be seen as the sum of two $2$D Fourier transforms in different orders, corresponding to the double-photon spectrum in frequency space. 

To normalize the second order correlation function in Eq. (\ref{eq57}) we next evaluate the number of photons coming out of the filter. The mean number of photons at the output can be written as,
\bea
\label{eq57}
\langle\hat{n}\rangle = v_g\int^{\infty}_{-\infty} d\omega~ \langle\tilde{\Psi}|\hat{E}^\dagger_{\text{out}}(\omega) \hat{E}_{\text{out}}(\omega) | \tilde{\Psi}\rangle.
\eea
On using Eqs. (\ref{eq53}- \ref{eq55}) we can write Eq. (\ref{eq57}) in terms of the spatially dependent field operators. Next on using the wave-function ansatz in Eq. (\ref{eq8}) and Eqs.  (\ref{eq31}-\ref{eq39}), we get
\bea
\label{eq58}
\langle \hat{n}\rangle & = &\frac{v_{g}}{2\pi} \int^{\infty}_{-\infty} d\omega\left|\mathcal{T}(\omega)\right|^{2}\bigg[\left|\int^{\zeta_T}_{0}d\zeta_{e}~\tilde{\phi}_{g\text{R}}(\zeta_{T},\zeta_{e})e^{-i\omega \zeta_{e}}\right|^{2}\nonumber\\
&+&\left|\int^{\zeta_T}_{0}d\zeta_{e}~\tilde{\phi}_{e\text{R}}(\zeta_{T},\zeta_{e})e^{-i\omega \zeta_{e}}\right|^{2}+\int^{\zeta_T}_{0}d\zeta_{e}\Bigg|\int^{\zeta_e}_{0}d\zeta'_{e}\nonumber\\
&\times&\tilde{\phi}_{\text{RR}}(\zeta_T,\zeta'_{e},\zeta_{e})e^{-i\omega \zeta'_{e}}+\int^{\zeta_T}_{\zeta_{e}}d\zeta'_{e}~\tilde{\phi}_{\text{RR}}(\zeta_T,\zeta_{e},\zeta'_{e})\nonumber\\
&\times&e^{-i\omega \zeta'_{e}}\Bigg|^{2}\bigg].
\eea

With the above derived expression, we simulate the normalized second order correlation function $g^{(2)}_{p} = \text{G}^{(2)}_{\text{pulsed}}/\langle n\rangle$. For this purpose we use the same Gaussian pulse defined in Sec. IV. A., combined with a Lorentzian filter with Transmission function $\mathcal{T}(\omega-\omega_{c}) = \left[\frac{\kappa/2}{i(\omega-\omega_{c})-\kappa/2}\right]$ or a Gaussian filter with Transmission function $\mathcal{T}(\omega-\omega_{c}) = \exp[-(\omega-\omega_{c})^2/2\kappa^{2}]$. Here $\kappa$ is the bandwidth of the filters. The accuracy of this simulation depends on the chosen time grid. The choice of step size is limited by the excitation pulse width, while the total time of the simulation should be large compared to the emitter's lifetime and the excitation pulse width. We choose a step size much smaller than the excitation pulse width keeping in mind that the inverse of the total time range should also be small compared to the bandwidth of the collection filter.
\begin{figure}
	\centering{}\includegraphics[scale=0.24]{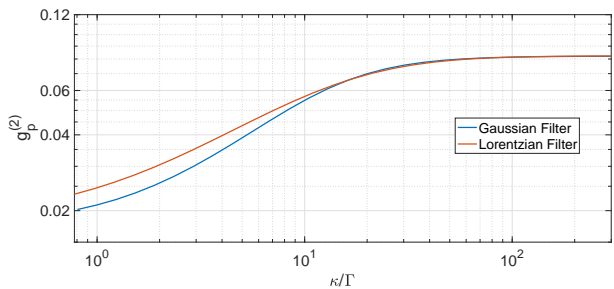}
	\caption{Comparison of spectral filtering of $g^{(2)}$ for different filters and a pulse width of $\sigma\Gamma = 0.1$ \label{fig8a}}
\end{figure}

In Fig. \ref{fig8} we show the effect of the filtering process on single photon transmission efficiency and $g^{(2)}$. While a filter can clean a single photon source, it has the drawback that it also filters out some of the desired single photons and thereby lowers the efficiency of the source. To study this behaviour we plot in  Fig. \ref{fig8} (a) and (b) the transmission efficiency of the filter for the single-photon component as a function of normalized filter bandwidth $\kappa/\Gamma$ for a Lorentzian and Gaussian filter respetively. The transmission of the filter is $\sim 50\%$ when the filter bandwidth is comparable to the inverse of the emitter lifetime. As the filter bandwidth increases to several times the decay rate, the transmission increases to $> 80\%$, which is more desirable for practical use. In Fig. \ref{fig8} (c) and (d), we plot $g^{(2)}_{p}$ as a function of the normalized excitation pulse width $\sigma\Gamma$ and filter bandwith $\kappa/\Gamma$ for Lorentzian and Gaussian filters respectively. We see from the figure that a narrow bandwidth filter can  strongly suppress $g^{(2)}_p$ by effectively filtering the two photon component in the emission. However as discussed above to achieve a reasonable transmission efficiency we need filter bandwidth of $\geq 10~\kappa/\Gamma$. In this regime, for both the filters we can achieve a low value $g^{(2)}_p$ for short pulses with durations $\sigma\Gamma \leq 0.1$ with only a minor reduction in the single photon efficiency. For large pulse durations $\sigma\Gamma \gtrsim 0.3$ though, to achieve a low $g^{(2)}_{p}$, a very narrow bandwidth $\kappa/\Gamma \lesssim1$ is required. In Fig. \ref{fig8a} we compare the effect of Gaussian and Lorentzian filters for an excitation pulse of duration $\sigma\Gamma =0.1$ on $g^{(2)}$. We find that for filter bandwidth $\leq 10~ \kappa/\Gamma$, the Gaussian filter works better than the Lorentzian filter in filtering the two photon components in emission and thereby lowering $g^{(2)}_{p}$ . This thus suggests that for a short excitation pulse using a Gaussian filter is benificial towards attaining a higher purity of a single photon source.   
 
\section{Steady state photon-photon correlation for emission from a system of emitters}
So far the treatment that we have presented is applicable for any time dependence of the transmitted field. We now focus on situations pertaining to constant incident field. In this case the dynamics approaches a steady state, but the solution to our equations will only be quasi-steady state because we only allow for a finite number of photons in the wavefunction, thus resulting in all population eventually ending up in the two photon emitted state. This means that eventually instead of reaching saturation, the dynamics of the states show a slow linear decay, which is proportional to a higher order of the input field $\mathcal{E}$. The issues arising from this apparent anomaly in the dynamics is discussed in detail in Ref. \cite{Mantas_thesis} considering all orders in the input-field . For this work, where we are mainly concerned with studying the normalized second order coherence, higher-order contributions and errors can be ignored by considering sufficiently low intensity input fields. 

\subsection{A single two level emitter}
To keep the analysis simple, we first discuss the steady state photon-photon correlation for emission from a single two level emitter coupled to a $1$D waveguide. This is a well studied problem (see for example \cite{Anders07}) and we revisit it here for pedagogical reasons. Once the dynamics attains the steady-state, the initial condition for photon-states (\ref{eq30}), becomes independent of the emission time. This means that to evaluate the dynamics of the system for all emission times, it is enough to consider a sufficiently large time $\zeta$, to ensure that the emitter state amplitudes $c_{g}$ and $c_{e}$ have attained steady-state. The photon-photon correlation function in Eq. (\ref{eq40}) then depends only on $\zeta$. This, significantly simplifies the problem and reduces the simulation time for evaluation of the second order coherence. 

 All the state amplitudes involved in the expression of $g^{(2)}$ function in Eq. (\ref{eq40}) can be evaluated by solving the set of equations Eqs. (\ref{eq31})-(\ref{eq39}) and (\ref{eq30})-(\ref{eq30a}). We solve the set of equations numerically and evaluate the corresponding second order correlation function $g^{(2)}(\zeta)$. Because of the issues above concerning our approach only being able to give quasi-steady state solutions care should be taken in how we evaluate expressions. In deriving Eqs. (\ref{eq40}) and (\ref{eqc2}) we have neglected all terms which are proportional to third and higher orders $\mathcal{O}(\mathcal{E}^{3})$ in the input field. The calculated results thus represent a consistent expansion of the correlation functions $\text{G}^{(1)}$ and $\text{G}^{(2)}$ up to second order in the intensity of the incoming field. 

In principle  we thus get correct results provided that we use a sufficiently low intensity. In practice, however, we find a better convergence by replacing the tems $\mathcal{E}^2$ in Eq. (\ref{eqc2}) by $|c_g|^2\mathcal{E}^2$ and $\mathcal{E}$ by $c_g \mathcal{E}$ in the first term in Eq. (\ref{eq40}). Up to second order in the intensity these expressions are the same, but because $|c_g|^2$ is reduced in the same manner as the single photon emission from  the emitter, this makes it easier to catch effects such a destructive interference in the forward direction giving rise to almost perfect reflection for strongly coupled emitters. Futhermore, in choosing the final time $T$ we have complete freedom with the only condition being that when all terms are included, for long enough time $T$  the intensity is independent of the time of measurement. For simulation purpose we have considered a final time of $T = 26/\Gamma$.

\begin{figure}
	\centering{}\includegraphics[scale=0.23]{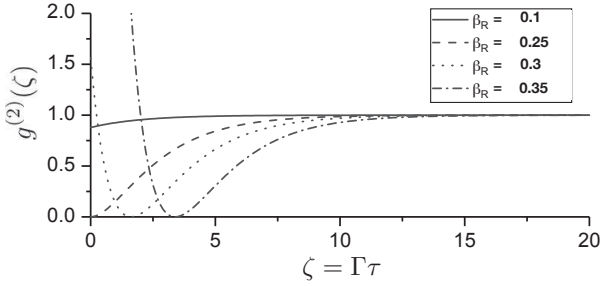}\caption{Second order correlation function $g^{(2)}(\zeta)$ for a single two-level quantum emitter at resonance ($\Delta=0$) with the incoming photon and coupled to a 1D waveguide. The properties of the emitted field for a single emitter only depends on $\beta_\text{R}$. \label{fig2}}
\end{figure}

The results of the simulation are presented in Fig. \ref{fig2} for different values of decay rates into the waveguide mode. When considering the transmission for a single emitter there is no difference between photons being emitted to the left and to the side. The transmission thus only depends on the value of $\beta_{\text{R}}$. Note that the analytical treatment of the second order correlation function that we discuss here is quite well known (see for example \cite{Chang07}) and we just reproduce it here for illustration and to lay the foundations for studying more complicated situations below. 

In Fig.  \ref{fig2} we see that for $\beta_{\text{R}}\leq0.25$, the $g^{(2)}$ function shows photon anti-bunching while for $\beta_{\text{R}}=0.25$ we find $g^{(2)}(0)=0$. These features are due to non-trivial interference effects between the emitted photons and the input field. This thus shows the complexity of the physics arising from even a  single scattering. Furthermore, for higher values of $\beta_{\text{R}}$, initial bunching occurs, which is followed by anti-bunching. Additionally, we see in Fig. \ref{fig2}, that,  $g^{(2)}(0)\sim1$ for low coupling factors, dropping to 0 at $\beta_{\text{R}} = 0.25$. For stronger couplings, one finds that $g^{(2)}(0)\gg1$, with a divergence at $\beta_{\text{R}}=0.5$, since in this case the transmission of the single-photon input state approaches zero.                                   
\subsection{Multiple two level emitters}
In this subsection we study the correlation characteristics of photons scattered from a system of multiple two level emitters coupled to a $1$D waveguide, by generalizing the wavefunction ansatz introduced in section III. To begin with, let us first write the generalized form of the wavefunction ansatz that was introduced in Eq. (\ref{eq25})
\begin{widetext}
	\bea
	\label{eq8}
	\Ket{\tilde{\Psi}(t)} &=&c_{g}(t)\Ket{g^{N}}\Ket{\emptyset}+\sum_{i=1}^{N}c_{e}^{i}(t)\Ket{e_{i}g^{N-1}}\Ket{\emptyset}+\sum_{i<j}^{N}c_{ee}^{ij}(t)\Ket{e_{i}e_{j}g^{N-2}}\Ket{\emptyset}+\int dt_{e}~\phi_{g\text{R}}(t,t_e)\hat{E}_{\text{R}}^{\dagger}(v_g(t-t_e))\Ket{g^{N}\emptyset}\nonumber\\
	&+&\sum_{m = 1}^{N}\int dt_{e}~\phi_{g\text{S},m}(t,t_e)\hat{E}_{\text{S},m}^{\dagger}(v_g(t-t_e))\Ket{g^{N}\emptyset}+\sum_{i=1}^{N}\int dt_e~\phi_{e\text{R}}^{i}(t,t_e)\hat{E}_{\text{R}}^{\dagger}(v_g(t-t_e))\Ket{e_{i}g^{N-1}\emptyset}\nonumber\\
	&+&\sum_{m = 1}^{N}\sum_{i=1}^{N}\int dt_e~\phi_{e\text{S},m}^{i}(t,t_e)\hat{E}_{\text{S},m}^{\dagger}(v_g(t-t_e))\Ket{e_{i}g^{N-1}\emptyset}+\int dt_{e2}\int dt_{e1}~\phi_{\text{RR}}(t,t_{e2},t_{e1})
	\hat{E}_{\text{R}}^{\dagger}(v_g(t-t_{e2}))\nonumber\\
	&\times&\hat{E}_{\text{R}}^{\dagger}(v_{g}(t-t_{e1}))\Ket{g^{N}\emptyset}+\int dt_{e2}\int dt_{e1}~\phi_{\text{RL}}(t,t_{e2},t_{e1})\hat{E}_{\text{R}}^{\dagger}(v_g(t-t_{e2}))\hat{E}_{\text{L}}^{\dagger}(v_g(t-t_{e1}))\Ket{g^{N}\emptyset}\nonumber\\
	&+&\sum_{m=1}^{N}\int dt_{e2}\int dt_{e1}~\phi_{\text{RS},m}(t,t_{e2},t_{e1})\hat{E}_{\text{R}}^{\dagger}(v_g(t-t_{e2}))\hat{E}_{\text{S},m}^{\dagger}(v_g(t-t_{e1}))\Ket{g^{N}\emptyset}+ \sum_{m=1}^{N}\int dt_{e2}\int dt_{e1}\nonumber\\
	&\times&\phi_{\text{SR},m}(t,t_{e2},t_{e1})\hat{E}_{\text{S},m}^{\dagger}(v_g(t-t_{e2}))\hat{E}_{\text{R}}^{\dagger}(v_g(t-t_{e1}))\Ket{g^{N}\emptyset}+\int dt_{e2}\int dt_{e1}\sum_{m=1}^{N}\sum_{m'= 1}^{N}~\phi_{\text{SS}mm',}(t,t_{e2},t_{e1})\nonumber\\
	&\times&\hat{E}_{\text{S},m}^{\dagger}(v_g(t-t_{e2}))\hat{E}_{\text{S},m'}^{\dagger}(v_g(t-t_{e1}))\Ket{g^{N}\emptyset}+\text{R}\leftrightarrow\text{L}.
	\eea
\end{widetext}
Here $c_{g}$ is the  amplitude of the combined ground state $|g^{N}\rangle = \prod_{j = 1}^{N}|g_{j}\rangle$ of all emitters with the field state being vacuum $|\emptyset\rangle$, while $c_{e}^{i}$ ($c_{ee}^{ij}$) is the amplitudes of the emitter excited state $\Ket{e_{i}g^{N-1}}(\Ket{e_{i}e_{j}g^{N-2}})$ with the $i^{th}$ emitter ($i^{th}$ and $j^{th}$ emitters) being excited while the other emitters being in the ground state, and the field state is in the vacuum state. The amplitude $\phi^{i}_{e\text{R(S,m)}}$ corresponds to the combined emitter-field state where the $i^{th}$ emitter is excited with all others in the ground state while an emitted photon is travelling (lost) to the right (side from emitter $m$). The other amplitudes bear the same interpretations as discussed for the single emitter wave-function ansatz.

We next derive the equation of motion for the various amplitudes in the above wave-function. For this purpose we use the Hamiltonian for the system of $N$-emitters given in Eq. (\ref{eq7}). As for the single emitter case, we can again divide the photon scattering process into two time domains, $0<t<t_{e}-\varepsilon$ and $t_{e}+\varepsilon < t < \infty$, corresponding to before and after emission of a photon respectively. Following the treatment in section III B  we can then write the dynamical equations of motions of the emitters state amplitudes in analogy to Eqs. (\ref{eq31}) - (\ref{eq32}) as
\bea
\label{eq9}
\dot{c}_{g}(\zeta) & = &i\sum_{j = 1}^{N}\frac{\Omega^\ast_j}{2}e^{i\tilde{\Delta}\zeta}c_{e}^{j}(\zeta),
\eea
\begin{widetext}
\bea
\label{eq10}
\dot{c}_{e}^{j}(\zeta) & = & i\frac{\Omega_{j}}{2}e^{-i\tilde{\Delta}\zeta}c_{g}(\zeta)-\frac{1}{2}c_{e}^{j}(\zeta)+i\sum_{l<j}\frac{\Omega^{\ast}_{l}}{2}e^{i\tilde{\Delta}\zeta}c_{ee}^{lj}(\zeta)+\sum_{j<l}\frac{\Omega^{\ast}_{l}}{2}e^{i\tilde{\Delta}\zeta}c_{ee}^{jl}(\zeta)-\beta_{\text{R}}\sum_{l<j}c_{e}^{j}(\zeta)e^{ik_{0}(\text{z}_{j}-\text{z}_{l})}\nonumber\\
&-&\beta_{\text{L}}\sum_{l>j}c_{e}^{l}(\zeta)e^{ik_{0}(\text{z}_{l}-\text{z}_{j})},\\
\label{eq11}
\dot{c}_{ee}^{jl}(\zeta) & = & i\frac{\Omega_l}{2}e^{-i\tilde{\Delta}\zeta}c_{e}^{j}(\zeta)+i\frac{\Omega_j}{2}e^{-i\tilde{\Delta}\zeta}c_{e}^{l}(\zeta)-\beta_{\text{R}}\bigg[\sum_{l'<j}c_{ee}^{l'j}(\zeta)e^{ik_{0}(\text{z}_{l}-\text{z}_{l'})}-\sum_{l' < l,l' > j}c_{ee}^{jl'}(\zeta)e^{ik_{0}(\text{z}_{l}-\text{z}_{l'})}-\sum_{j'<j}c_{ee}^{j'l}(\zeta)\nonumber\\
&\times&e^{ik_{0}(\text{z}_{j}-\text{z}_{j'})}\bigg]-\beta_{\text{L}}\bigg[\sum_{l'>l}c_{\text{ee}}^{jl'}(\zeta)e^{ik_{0}(\text{z}_{l'}-\text{z}_{l})}-\sum_{j'>j,j'<l}^{N}c_{ee}^{j'l}(\zeta)e^{ik_{0}(\text{z}_{j'}-\text{z}_{j})}-\sum_{j'>l}^{N}c_{ee}^{lj'}(\zeta)e^{ik_{0}(\text{z}_{j'}-\text{z}_{j})}\bigg],
\eea
\end{widetext}
where as before we have used a dimensionless time variables $(\zeta, \zeta_{e})$. Note that the above set of equations give the dynamics of the emitters before the emission of a photon i.e. during the time domain (in dimensionless co-ordinate) $0<\zeta <\zeta_{e}+\varepsilon $. 

From the above set of Eqs. (\ref{eq9}-\ref{eq11}) we see that similar to the single emitter case, the ground state amplitude only changes due to the excitation process of the emitters by the input field. The amplitude representing the single excitation however in comparison to the single emitter case, now involves additional terms due to the presence of other emitters. Given that the emitters are spatially distributed in the $1$D waveguide, a photon emitted by one emitter can propagate along the waveguide and excite another emitter at some other location at a later time thereby coupling the excited state amplitudes of the two emitters. This is represented by terms in Eq. (\ref{eq10}) where the photon acquires a phase factor $\mathrm{e}^{ik_{0}(\text{z}_{l}-\text{z}_{j})}$ that is proportional to the path travelled by the photon between the emitters. As such, the evolution of the excited state amplitude of any $j^{th}$ emitter gets contribution from all possible scattering events for photons travelling to the right or left of the $j^{th}$ emitter. Furthermore, the single excitation amplitude is also coupled to the amplitude corresponding to more than one emitter being excited, represented by the $c^{ij}_{ee}$ term.  The dynamics of the double excitation process on the other hand is more complicated. The dynamics in the double excitation subspace reflects the same processes, only now the terms look more complicated since the absorption/emission dynamics involve both indices and we have made the convention that $c^{jl}_{ee}$ vanishes unless $j<l$. Note that, the coupling between the emitters leads to quantum interference of the excited state amplitudes and creates interaction between the emitted photons that is manifested in the photon correlation characteristics.

For the other time domain $(\zeta_{e}+\epsilon < \zeta < \infty)$ after emission of a photon, in analogy to Eqs. (\ref{eq38})- (\ref{eq39}) of section III B, we find the equations of motion of the amplitudes of the field-emitter states as 
\bea
\label{eq14}
\dot{\tilde{\phi}}_{g\text{R}}(\zeta,\zeta_{e}) & = & i\sum_{j=1}^{N}\frac{\Omega^\ast_{j}}{2}e^{i\tilde{\Delta}\zeta}\tilde{\phi}_{e\text{R}}^{j}(\zeta,\zeta_{e}),\\
\label{eq15}
\dot{\tilde{\phi}}_{e\text{R}}^{j}(\zeta,\zeta_{e})& = & i\frac{\Omega_{j}}{2}e^{-i\tilde{\Delta}\zeta}\tilde{\phi}_{g\text{R}}(\zeta,\zeta_{e})-\frac{1}{2}\tilde{\phi}_{e\text{R}}^{j}(\zeta,\zeta_{e})\nonumber\\
&-&\beta_{\text{R}}\sum_{l<j}\tilde{\phi}_{e\text{R}}^{l}(\zeta,\zeta_{e})e^{ik_{0}(\text{z}_{j}-\text{z}_{l})}-\beta_{\text{L}}\sum_{l>j}\tilde{\phi}_{e\text{R}}^{j}(\zeta,\zeta_{e})\nonumber\\
&\times&e^{ik_{0}(\text{z}_{l}-\text{z}_{j})}.
\eea
\begin{figure}[!h]
	\centering{}\includegraphics[scale=0.5]{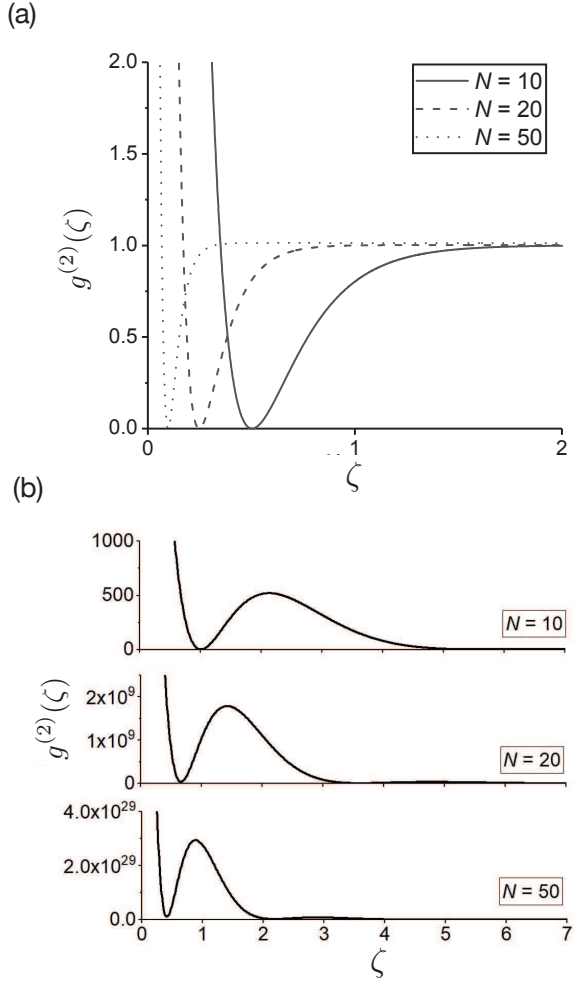}\caption{Second order correlation function $g^{(2)}(\zeta)$ for photons emitted from $N$ two-level emitters on resonance ($\Delta=0$) with the incoming photon and coupled to a 1D waveguide. (a) $\beta_{\text{R}} = \beta_{\text{L}} = 0.45$ and $k_{0}\Delta\text{z} = \pi$. (b) $\beta_{\text{R}} = \beta_{\text{L}} = 0.15$ and $k\Delta\text{z} = \pi/4$. \label{fig4}}
\end{figure}
Note that there exists an equivalent set of equations for the left going photons and photons going out of the waveguide given in appendix B. We here however, choose the right propagating direction as the preferred one for measurement and hence will restrict our studies to the dynamics of the right going photons.

In regards to the amplitudes in Eq. (\ref{eq14}) and Eq. (\ref{eq15}) we find that the equation of the propagating photons is similar to that governing the emitter dynamics. The only difference is that we here exclude doubly excited states, since such a state plus a photon propagating to the right would constitute a triply excited states. In defining our wave-function ansatz we have already neglected such a state by truncating the photonic part of the wave-function to two excitations in the weak field approximation. This thus just reflect that the photon does not affect the dynamics once it has left the system. Since the photon can leave at different times these equations of motion, however,  should be solved for each emission time $\zeta_e$ subject to the initial conditions. 

In the following we will use the solution of these two sets of equations Eq. (\ref{eq9}) - Eq. (\ref{eq11}) and Eq. (\ref{eq14}) - Eq. (\ref{eq15}) to study the correlation characteristics of two photon emission from a system of $N$-emitters. To solve the above set of equations, similar to Eq. (\ref{eq30}) of section III B we invoke the following initial conditions 
\bea
\label{eq16}
\tilde{\phi}_{g\text{R}}(\zeta_{e}+\varepsilon,\zeta_{e})   & = &  i\sqrt{\beta_{\text{R}}}\sum_{j=1}^{N}~c_{e}^{j}(\zeta_{e})e^{i\tilde{\Delta}\zeta_{e}-ik_{0}\text{z}_{j}},\\
\label{eq17}
\tilde{\phi}^{j}_{e\text{R}}(\zeta_{e}+\varepsilon,\zeta_{e}) & = & i\sqrt{\beta_{\text{R}}}\sum_{l<j}^{N}c_{ee}^{jl}(\zeta_{e})e^{i\tilde{\Delta}\zeta_{e}-ik_{0}\text{z}_{l}}\nonumber\\
&+&i\sqrt{\beta_{\text{R}}}\sum_{l>j}^{N}c_{ee}^{lj}(\zeta_{e})e^{i\tilde{\Delta}\zeta_{\mathrm{e}}-ik_{0}\text{z}_{l}}.\\
\dot{\tilde{\phi}}_{\text{pq}}(\zeta_{e2}+\varepsilon,\zeta_{e2},\zeta_{e1}) & = & i\sqrt{\beta_{\text{q}}}\sum_{j=1}^{N}\tilde{\phi}_{e\text{q}}^{j}(\zeta_{e2},\zeta_{e1})e^{i\tilde{\Delta}\zeta_{e2}-ik_{0}\text{z}_{e_j}},\nonumber\\
\eea
with $\{\text{p,q}\} =\{\text{R}, \text{L}, \text{S} \}$. The initial conditions for the single photon amplitudes $\tilde{\phi}_{(e/g)\text{p}}$ express that the amplitude to emit a photon at time $\zeta_{e}$ is proportional to the amplitude of being excited at that time. The initial condition for the two photon amplitudes $\tilde{\phi}_{\text{pq}}$ similarly express that amplitude to emit a second photon at time $\zeta_{e2}$ is proportional to the single photon amplitude at time $\zeta_{e1}$ and having an excitation in the system at $\zeta_{e2}$. In the wave-function ansatz of Eq. (\ref{eq8}), the components representing the photons in principle should have amplitudes with explicit position dependence, to account for photons emitted at different positions along the waveguide. For most practical situations, however, the size of the ensemble is sufficiently small that one can neglect any retardation effects due to the time it takes for a photon to travel through the ensemble (apart from the phase factor $e^{ik_0\text{z}}$ already accounted for in Eq. (\ref{eq5})). We have therefore ignored all effects of retardation here.  

Following the definition of the second order correlation function from section III C we next evaluate the normalized second order correlation function $g^{(2)}$ and investigate the correlation properties of photons scattered from such $N$ emitter systems. We show in Fig. \ref{fig4} the behaviour of the second order correlation function $g^{(2)}$ for $N$ emitters. In the $N$-emitter case, however, one can see from Fig. \ref{fig4} (a) that, for emitters spatially separated by $k_{0}\Delta\text{z} = \pi$ there is strong bunching between emitted photons near zero delay. This is because there is constructive interference for reflection leading to a very low transmission of single photons. On the other hand when two photons are incident simultaneously they can exchange energy, such that one is red shifted and the other is blue shifted. Neither of the photons are thus resonant with the emitters and will have an increased transmission. Furthermore, there is a sign change of the two photon wavefunction $\tilde{\phi}_{\text{RR}}(\zeta, \zeta_{e2}, \zeta_{e1})$ at the same time coordinate $\zeta_{e1}=\zeta_{e2}$ compared to when they are different $\zeta_{e2} - \zeta_{e1} \gg 1 $. This leads to a node in $\tilde{\phi}_{\text{RR}}(\zeta, \zeta_{e2}, \zeta_{e1})$ corresponding to $g^{(2)}(\zeta_0) = 0$ at the position $\zeta_0=\zeta_{e2}-\zeta_{e1}$, where the wavefunction change sign \cite{Sahand18}. For $k_{0} \Delta\text{z} = \pi/4$ and strong coupling $(\beta_\text{R} = \beta_\text{L} = 0.45)$ we show the dynamical behaviour of $g^{(2)}$ in Fig. \ref{fig4} (b). We see strong bunching along with rapid amplitude oscillations in $g^{(2)}$. This is due to multiple scattering of the photons in $N$ emitter systems. In general there are several different physical effects which can be explored with this method for the multi-emitter system. A full investigation of this is beyond the scope of the present article. We refer the reader to \cite{Sahand18} where this method was used to investigate some of the effects attainable in such multi-emitter system. 
\section{Conclusion}
In conclusion, we have introduced a formalism based on the time-dependent wave-function ansatz, to calculate multi-time correlations of photons emitted from a system of $N$ quantum emitters weakly coupled to a $1$D waveguide. Our formalism is valid in the limit of small total number of emitted photons and is advantageous compared to methods based on the quantum regression theorem for large Hilbert spaces. For simplicity we have restricted the discussion to two level emitters and have assumed all of them to be identical. It is, however, straightforward to generalize the formalism to other level structures or emitters having different properties.

We have applied our formalism to model the spectral filtering of photons emitted by a quantum emitter coupled to a $1$D waveguide. Compared to the quantum regression theorem our wave function approach is much more suited for such spectral filtering problems as our method involves only calculating the two-photon component $\phi_{\text{RR}}(\zeta_T,\zeta_e',\zeta_e)$ of our wave function with a single $\zeta_T$. Furthermore we have discussed how our formalism can be used to study photon dynamics in quantum many-body systems. An application of this formalism in this regards can be found in Ref. \cite{Sahand18}.

Our formalism has strong similarities to the Monte Carlo wave function technique, with the additional feature that it gives an explicit description of the outgoing quantum state. This is advantageous when the scattered photons are subject to further evolution and not detected right away. An example of this is the filtered single photon source that we studied in this work. However, the Monte Carlo method has the advantage of Monte Carlo sampling multiple decays to various reservoirs, whereas our method needs to keep track of each individual decay to each reservoir. For dynamics with multiple decays to several reservoirs, our fomalism quickly becomes intractable and  it is preferable to use the Monte Carlo approach. It would be desirable to combine our technique of a wavefunction ansatz for the forward propagating direction with Monte Carlo sampling of all unobserved degrees of freedom \cite{Molmer19}. In this way one could achieve an efficient description of the forward propagation including, e.g. frequency filtering while simultaneously having less restrictions on the total number of emitted photons. 
 
\begin{acknowledgments}
SD, PL, and AS acknowledge financial support from the Danish Council for Independent Research - Natural Sciences
(FNU), and the Danish National Research Foundation (Center of Excellence "Hy-Q", Grant No. DNRF139), LZ received funding from the European Union Horizon 2020 Research and Innovation programme under the Marie Sk\l{}odowska Curie grant agreement No. 721394 (4PHOTON).
\end{acknowledgments} 
\appendix
\begin{widetext}

\section{Details of the Equations of Motion in Sec. III for Single Emitter System}
Reducing the Hamiltonian in Eq. (\ref{eq7}) to the single emitter case, expanding the Schr\"{o}dinger equation $|\dot{\tilde{\Psi}}\rangle = \left(-i/\hbar\right)\hat{\tilde{\mathcal{H}}}|\tilde{\Psi}\rangle$ and using Eq. (\ref{eq25}) we get the equation for the state amplitudes for the emitters in the time domain $0<\zeta<\zeta_{\mathrm{e}}-\varepsilon$ as 
\bea
\label{eqb1}
\dot{c}_{g}(\zeta) &= &i\sqrt{\beta_{\text{R}}}\mathcal{{\tilde{E}}}^{*}e^{i\tilde{\Delta}\zeta}c_{e}(\zeta),\\
\label{eqb2}
\dot{c}_{e}(\zeta) & = &i\sqrt{\beta_{\text{R}}}\mathcal{{\tilde{E}}}e^{-i\tilde{\Delta}\zeta}c_{g}(\zeta)+i\sum_{j}\sqrt{\beta_{j}}\tilde{\phi}_{g\text{j}}(\zeta,\zeta)e^{-i\tilde{\Delta}\zeta},\\
\label{eqb3}
\dot{\tilde{\phi}}_{gj}(\zeta,\zeta_{e}) & = & i\sqrt{\beta_{j}}c_e(\zeta)e^{i\tilde{\Delta}\zeta}\delta(\zeta - \zeta_{e}),
\eea
while the one corresponding to after the emission of a photon in the time domain $\zeta_{\mathrm{e}}+\varepsilon<\zeta<\infty$ are 
\bea
\label{eqb4}
\dot{\tilde{\phi}}_{gj}(\zeta,\zeta_{e}) & = & i\sqrt{\beta_{j}}\mathcal{{\tilde{E}}}^{*}e^{i\tilde{\Delta}\zeta}\tilde{\phi}_{ej}(\zeta,\zeta_{e}),\\
\label{eqb5}
\dot{\tilde{\phi}}_{ej}(\zeta,\zeta_{e})& = & i\sqrt{\beta_{j}}\mathcal{{\tilde{E}}}e^{-i\tilde{\Delta}\zeta}\tilde{\phi}_{gj}(\zeta,\zeta_{e})-\frac{\beta_{\text{s}}}{2}\tilde{\phi}_{ej}(\zeta,\zeta_{e})+i\sum_{k}\sqrt{\beta_{k}}\tilde{\phi}_{jk}(\zeta,\zeta,\zeta_{e})e^{-i\tilde{\Delta}\zeta}
\eea
Here $j, k$ corresponds to the direction of the photon propagation $\{\text{R},\text{L}, \text{S}\}$. Note that in deriving the above set of equations we have used as before, dimensionless variables. In the next appendix, we show the generalization of the approach to multiple emitters, and explicitly write down the equations of motions of the amplitudes for photon lost to side (out of the waveguide) to do a detail discussion of the dynamics. 

We see from Eq. (\ref{eqb3}), that it only depends on the amplitude of the excited state and it can be formally integrated with respect to $\zeta$ to give 
\bea
\label{eqb6}
\tilde{\phi}_{g\text{j}}(\zeta_{e}+\varepsilon,\zeta_e) & = & i\sqrt{\beta_{j}}c_e(\zeta_{e})e^{i\tilde{\Delta}\zeta_{e}},
\eea
Substituting this for $\tilde{\phi}_{g\text{R}}$ and $\tilde{\phi}_{g\text{L}}$ in Eq. (\ref{eqb2}) we get , the evolution of the system before the emission to be governed by 
\bea
\label{eqb7}
\dot{c}_{g}(\zeta) & = &i\sqrt{\beta_{\mathrm{R}}}\mathcal{{\tilde{E}}}^{*}e^{i\tilde{\Delta}\zeta}c_e(\zeta),\\
\label{eqb8}
\dot{c}_{e}(\zeta) & = &i\sqrt{\beta_{\mathrm{R}}}\mathcal{{\tilde{E}}}e^{-i\tilde{\Delta}\zeta}c_{g}(\zeta)-\frac{1}{2}c_e(\zeta).
\eea
Here we have used $\theta(0) = \frac{1}{2}$, which is consistent with standard results obtained in the Markow approximation (excluding the Lamb shift), and we have $\beta_{\text{R}}+\beta_{\text{L}}+\beta_{\text{s}} = 1$. The dynamics of the two-level emitter can then be simply interpreted as follows. There are two processes that affect the dynamical evolution of emitter states until the emission of a photon: $(1)$ excitation of the ground state by the input field (the terms proportional to $\mathcal{{E}}$ in Eq (\ref{eqb7})), and $(2)$ the decay of the excited state into any channel (given by $-\frac{1}{2}c_e$).

A similar derivation for the time domain $\zeta_{e}+\varepsilon<\zeta<\infty$ gives us the dynamical evolution of the amplitudes corresponding to the two-photon components of the wave-function ansatz 
\bea
\label{eqb10}
\dot{\tilde{\phi}}_{jk}(\zeta,\zeta_{e2},\zeta_{e1}) & = & i\sqrt{\beta_\text{j}}\tilde{\phi}_{ek}(\zeta,\zeta_{e1})e^{i\tilde{\Delta}\zeta}\delta(\zeta - \zeta_{e2}).
\eea
Here as before $j$ and $k$ both correspond to the direction of photon propagation given by $\{\text{R, L, S}\}$. Integrating the terms in Eq. (\ref{eqb10}) that contributes to the two-photon component then gives
\bea
\label{eqb12}
\tilde{\phi}_{jk}(\zeta_{e_2}+\varepsilon,\zeta_{e2},\zeta_{e1}) & = & i\sqrt{\beta_\text{j}}\tilde{\phi}_{ek}(\zeta_{e2},\zeta_{e1})e^{i\tilde{\Delta}\zeta_{e2}}.
\eea

Next using Eq (\ref{eqb12}) and substituting for $\tilde{\phi}_{\text{RR}}, \tilde{\phi}_{\text{SR}}, \tilde{\phi}_{\text{LR}}$ in Eq. (\ref{eqb5}) when $j = \text{R}$, we get the dynamical evolution of the amplitude $\tilde{\phi}_{e\text{R}}$ as
\bea
\label{eqb13}
\dot{\tilde{\phi}}_{e\text{R}}(\zeta,\zeta_{e})& = &i\sqrt{\beta_{\text{R}}}\mathcal{{\tilde{E}}}\mathrm{e}^{-i\tilde{\Delta}\zeta}\tilde{\phi}_{g\text{R}}(\zeta,\zeta_{e})-\frac{1}{2}\tilde{\phi}_{e\text{R}}(\zeta,\zeta_{e}).
\eea
Note that in deriving Eq. (\ref{eqb13}) we have considered the contribution of the two photon components $\tilde{\phi}_{\text{RR}}(\zeta,\zeta_{e},\zeta) = \tilde{\phi}_{\text{RL}}(\zeta,\zeta_{e},\zeta) = 0$. This can be seen easily from Eq (\ref{eqb12}), as the amplitudes $\tilde{\phi}_{e\text{R}}(\zeta,\zeta) = \tilde{\phi}_{e\text{L}}(\zeta,\zeta) = 0$, following the initial condition that at the emission time these state amplitudes are zero. So from Eqs. (\ref{eqb4}) and (\ref{eqb5}) we find that the time-evolution of a right going photon after emission is described by the following system of equations:
\bea
\label{eqb14}
\dot{\tilde{\phi}}_{g\text{R}}(\zeta,\zeta_{e})& = &i\sqrt{\beta_{\text{R}}}\mathcal{{\tilde{E}}}^{*}e^{i\tilde{\Delta}\zeta}\tilde{\phi}_{e\text{R}}(\zeta,\zeta_{e}),\\
\label{eqb15}
\dot{\tilde{\phi}}_{e\text{R}}(\zeta,\zeta_{e})& = &i\sqrt{\beta_{\text{R}}}\mathcal{{\tilde{E}}}e^{-i\tilde{\Delta}\zeta}\tilde{\phi}_{g\text{R}}(\zeta,\zeta_{e})-\frac{1}{2}\tilde{\phi}_{e\text{R}}(\zeta,\zeta_{e}).
\eea
The above equations are similar to the system of equations representing the dynamics of the emitter states in Eqs. (\ref{eqb7}) and (\ref{eqb8}). This is expected for scattering from a single emitter because, once the emitter decays and emits a photon, the evolution of the states and the photon behaves in the same way. Note that, the initial condition for solving the system of Eqs. (\ref{eqb14}) and (\ref{eqb15}) is given by Eq. (\ref{eqb6}).

Note that all physical variables in the system are normalized with respect to $\Gamma$, giving new dimensionless variables in the form $\mathcal{{\tilde{E}}} = \Gamma \mathcal{{E}}/v_{g}$, $\tilde{\phi}_{g\text{R(L)}} = \phi_{g\text{R(L)}}/\sqrt{\Gamma v_{g}}$, $\tilde{\phi}_{e\text{R(L)}} = \phi_{e\text{R(L)}}/\sqrt{\Gamma v_{g}}$, $\tilde{\phi}_{\text{R(L)R(L)}} = \phi_{\text{R(L)R(L)}}/\Gamma v_{g}$, $\tilde{\Delta} = \Delta/\Gamma$, $\beta_{\mathrm{R(L)}} = \mathrm{\Gamma_{1D,R(L)}}/\Gamma$ and, $\beta_{s} = \Gamma'/\Gamma$, satisfying the normalization condition $\beta_{\text{R}}+\beta_{\text{L}}+\beta_{s}=1$.

\section{Details of the Equations of Motion in Sec. VI for a Multiple Emitter System}
Expanding the Schr\"{o}dinger equation (\ref{eq7}) in the form $|\dot{\tilde{\Psi}}\rangle = \left(-i/\hbar\right)\hat{\tilde{\mathcal{H}}}|\tilde{\Psi}\rangle$ and using Eq. (\ref{eq8}) we get the equation for the state amplitudes for the emitters 
\bea
\label{a4}
\dot{c}_{g(\zeta)} & = & i\sqrt{\beta_{\text{R}}}\sum_{i=1}^{N}\tilde{\mathcal{{E}}}^{*}e^{i\tilde{\Delta}\zeta-ik_{0}\text{z}_{i}}c_{e}^{i}(\zeta),\\
\label{a5}
\dot{c}_{e}^{j}(\zeta) & = & i\sqrt{\beta_{\text{R}}}\tilde{\mathcal{{E}}}e^{-i\tilde{\Delta}\zeta+ik_{0}\text{z}_{i}}c_{g}(\zeta)+i\sqrt{\beta_{s}}\tilde{\phi}^{j}_{g\text{S}}(\zeta,\zeta)e^{-i\tilde{\Delta}\zeta}+i\sqrt{\beta_{\text{R}}}\sum_{l<j}^{N}\tilde{\mathcal{{E}}}^{*}e^{i\tilde{\Delta}\zeta-ik_{0}\text{z}_{j}}c_{ee}^{lj}(\zeta)+ i\sqrt{\beta_{\text{R}}}\sum_{j<l}^{N}\tilde{\mathcal{{E}}}^{*}e^{i\tilde{\Delta}\zeta-ik_{0}\text{z}_{j}}c_{ee}^{jl}(\zeta)\nonumber\\
&+ &i\sqrt{\beta_{\text{R}}} \tilde{\phi}_{g\text{R}}(\zeta,\zeta-\Gamma\frac{\text{z}_{i}}{v_{g}})e^{-i\tilde{\Delta}\zeta+ik_{0}\text{z}_{i}}+ i\sqrt{\beta_{\text{R}}}\tilde{\phi}_{g\text{L}}(\zeta,\zeta+\Gamma\frac{\text{z}_{i}}{v_{g}})e^{-i\tilde{\Delta}\zeta-ik_{0}\text{z}_{i}},\\
\label{a6}
\dot{c}_{ee}^{jl}(\zeta) & = & i\sqrt{\beta_{\text{R}}}\tilde{\mathcal{{E}}}e^{-i\tilde{\Delta}\zeta+ik_{0}\text{z}_{j}}c_{e}^{j}(\zeta)+i\sqrt{\beta_{\text{R}}}\tilde{\mathcal{{E}}}e^{-i\tilde{\Delta}\zeta+ik_{0}\text{z}_{i}}c_{e}^{l}(\zeta)+i\sqrt{\beta_{s}}\tilde{\phi}^{j}_{e\text{S}}(\zeta,\zeta)e^{-i\tilde{\Delta}\zeta}+i\sqrt{\beta_{s}}\tilde{\phi}^{l}_{e\text{S}}(\zeta,\zeta)e^{-i\tilde{\Delta}\zeta}\nonumber\\ &+&i\sqrt{\beta_{\text{R}}}\tilde{\phi}_{e\text{R}}^{j}(\zeta,\zeta-\Gamma\frac{\text{z}_{l}}{v_{g}})e^{-i\tilde{\Delta}\zeta+ik_{0}\text{z}_{l}}+ i\sqrt{\beta_{\text{R}}}\tilde{\phi}_{e\text{R}}^{l}(\zeta,\zeta-\Gamma\frac{\text{z}_{j}}{v_{g}})e^{-i\tilde{\Delta}\zeta+ik_{0}\text{z}_{j}}\nonumber\\ &+&i\sqrt{\beta_{\text{L}}}\tilde{\phi}_{e\text{L}}^{j}(\zeta,\zeta+\Gamma\frac{\text{z}_{l}}{v_{g}})e^{-i\tilde{\Delta}\zeta-ik_{0}\text{z}_{l}}+ i\sqrt{\beta_{\text{L}}}\tilde{\phi}_{e\text{L}}^{l}(\zeta,\zeta+\Gamma\frac{\text{z}_{j}}{v_{g}})e^{-i\tilde{\Delta}\zeta-ik_{0}\text{z}_{j}}.
\eea
In writing the above set of equations we have used the dimensionless variable $\zeta = \Gamma t$. We get similar equations of motion for the amplitudes of the photonic part of the wave-function
\bea
\label{a7}
\dot{\tilde{\phi}}_{g\text{R}}(\zeta,\zeta_{e}) &  = &  i\sqrt{\beta_{\text{R}}}\sum_{j=1}^{N}c_{e}^{j}(\zeta_{e})\delta(\zeta_{e}-\zeta_{e})e^{i\tilde{\Delta}\zeta_{e}-ik_{0}\text{z}_{e}}+ i\sqrt{\beta_{\text{R}}}\sum_{j=1}^{N}\tilde{\mathcal{E}}^{*}e^{i\tilde{\Delta}\zeta-ik_{0}\text{z}_{j}}\phi_{e\text{R}}^{j}(\zeta,\zeta_{e}),\\
\label{a7a}
\dot{\tilde{\phi}}_{e\text{R}}^{j}(\zeta,\zeta_{e}) & = & i\sqrt{\beta_{\text{R}}}\sum_{l<j}^{N}c_{ee}^{lj}(\zeta_{e})\delta(\zeta-\zeta_{e})e^{i\tilde{\Delta}\zeta_{\mathrm{e}}-ik_{0}\text{z}_{e}}+i\sqrt{\beta_{\text{R}}}\sum_{l>j}^{N}c_{ee}^{jl}(\zeta_{e})\delta(\zeta-\zeta_{e})e^{i\tilde{\Delta}\zeta_{e}-ik_{0}\text{z}_{e}}\nonumber\\
& + & i\sqrt{\beta_{\text{R}}}\tilde{\mathcal{E}}e^{-i\tilde{\Delta}\zeta+ik_{0}\text{z}_{j}}\tilde{\phi}_{g\text{R}}(\zeta,\zeta_{e})-\frac{\beta_{s}}{2}\tilde{\phi}_{e\text{R}}^{j}(\zeta,\zeta_{e})+i\sqrt{\beta_{\text{R}}}\tilde{\phi}_{\text{RR}}(\zeta,\zeta-\Gamma\frac{\text{z}_{j}}{v_{g}},\zeta_{e})\nonumber\\
& \times&e^{-i\tilde{\Delta}\zeta+ik_{0}\text{z}_{j}}+ i\sqrt{\beta_{\text{R}}}\tilde{\phi}_{\text{RR}}(\zeta,\zeta_{e},\zeta-\Gamma\frac{\text{z}_{j}}{v_{g}})e^{-i\tilde{\Delta}\zeta+ik_{0}\text{z}_{j}}+ i\sqrt{\beta_{\text{L}}}\nonumber\\
&\times &\tilde{\phi}_{\text{LR}}(\zeta,\zeta+\Gamma\frac{\text{z}_{j}}{v_{g}})e^{-i\tilde{\Delta}\zeta-ik_{0}\text{z}_{j}}+ i\sqrt{\beta_{\text{L}}}\tilde{\phi}_{\text{RL}}(\zeta,\zeta_{e},\zeta+\Gamma\frac{\text{z}_{j}}{v_{g}})e^{-i\tilde{\Delta}\zeta-ik_{0}\text{z}_{j}},\\
\label{a8}
\dot{\tilde{\phi}}_{g\text{L}}(\zeta,\zeta_{e}) & = & i\sqrt{\beta_{\text{L}}}\sum_{j=1}^{N}c_{e}^{i}(\zeta_{e})\delta(\zeta-\zeta_{e})e^{i\tilde{\Delta}\zeta_{e}+ik_{0}\text{z}_{e}}+i\sqrt{\beta_{\text{R}}}\sum_{j=1}^{N}\tilde{\mathcal{E}}^{*}e^{i\tilde{\Delta}\zeta-ik_{0}\text{z}_{j}}\tilde{\phi}_{e\text{L}}^{j}(\zeta,\zeta_{e}),\\
\label{a9}
\dot{\tilde{\phi}}_{e\text{L}}^{j}(\zeta,\zeta_{e}) & = & i\sqrt{\beta_{\text{L}}}\sum_{l<j}^{N}c_{ee}^{lj}(\zeta_{e})\delta(\zeta-\zeta_{e})e^{i\tilde{\Delta}\zeta_{e}+ik_{0}\text{z}_{e}}+i\sqrt{\beta_{\text{L}}}\sum_{l>j}^{N}c_{ee}^{jl}(\zeta_{e})\delta(\zeta-\zeta_{e})e^{i\tilde{\Delta}\zeta_{e}+ik_{0}\text{z}_{e}}\nonumber\\
&+& i\sqrt{\beta_{\text{R}}}\tilde{\mathcal{E}}e^{-i\tilde{\Delta}\zeta+ik_{0}\text{z}_{j}}\tilde{\phi}_{g\text{L}}(\zeta,\zeta_{e})-\frac{\beta_{s}}{2}\tilde{\phi}_{e\text{L}}^{j}(\zeta,\zeta_{e})+i\sqrt{\beta_{\text{L}}}\tilde{\phi}_{\text{LL}}(\zeta,\zeta-\Gamma\frac{\text{z}_{j}}{v_{g}},\zeta_{e})\nonumber\\
&\times&e^{-i\tilde{\Delta}\zeta-ik_{0}\text{z}_{j}}+ i\sqrt{\beta_{\text{L}}}\tilde{\phi}_{\text{LL}}(\zeta,\zeta_{e},\zeta-\Gamma\frac{\text{z}_{j}}{v_{g}})e^{-i\tilde{\Delta}\zeta-ik_{0}\text{z}_{j}}+i\sqrt{\beta_{\text{R}}}\nonumber\\
&\times&\tilde{\phi}_{\text{RL}}(\zeta,\zeta_{e},\zeta+\Gamma\frac{\text{z}_{j}}{v_{g}})e^{-i\tilde{\Delta}\zeta+ik_{0}\text{z}_{j}}+ i\sqrt{\beta_{\text{R}}}\tilde{\phi}_{\text{LR}}(\zeta,\zeta+\Gamma\frac{\text{z}_{j}}{v_{g}},\zeta_{e})e^{-i\tilde{\Delta}\zeta+ik_{0}\text{z}_{j}}.
\eea
and the two photon amplitudes as
\bea
\label{a10}
\dot{\tilde{\phi}}_{\text{RR}}(\zeta,\zeta_{e2},\zeta_{e1}) & = & i\sqrt{\beta_{\text{R}}}\sum_{j=1}^{N}\tilde{\phi}_{e\text{R}}^{j}(\zeta_{e2},\zeta_{e1})\delta(\zeta-\zeta_{e2})e^{i\tilde{\Delta}\zeta_{e2}-ik_{0}\text{z}_{e_j}},\\
\label{a11}
\dot{\tilde{\phi}}_{\text{LR}}(\zeta,\zeta_{e2},\zeta_{e1}) & = & i\sqrt{\beta_{\text{L}}}\sum_{j=1}^{N}\tilde{\phi}_{e\text{R}}^{j}(\zeta_{e2},\zeta_{e1})\delta(\zeta-\zeta_{e2})e^{i\tilde{\Delta}\zeta_{e2}+ik_{0}\text{z}_{e_j}},\\
\label{a12}
 \dot{\tilde{\phi}}_{\text{RL}}(\zeta,\zeta_{e2},\zeta_{e1}) & = & i\sqrt{\beta_{\text{R}}}\sum_{j=1}^{N}\tilde{\phi}_{e\text{L}}^{j}(\zeta_{e2},\zeta_{e1})\delta(\zeta-\zeta_{e2})e^{i\tilde{\Delta}\zeta_{e2}-ik_{0}\text{z}_{e_j}},\\
 \label{a13}
 \dot{\tilde{\phi}}_{\text{LL}}(\zeta,\zeta_{e2},\zeta_{e1}) & = & i\sqrt{\beta_{\text{L}}}\sum_{j=1}^{N}\tilde{\phi}_{e\text{L}}^{j}(\zeta_{e2},\zeta_{e1})\delta(\zeta-\zeta_{e2})e^{i\tilde{\Delta}\zeta_{e2}+ik_{0}\text{z}_{e_j}}.
\eea
A similar set of equations exits for the photons lost to the outside. We list them here below,
\bea
\dot{\tilde{\phi}}_{\mathrm{gS}m}(\zeta,\zeta_{e})& = & i\sqrt{\beta_{s}}\sum_{j = 1}^{N}c^{j}_{\mathrm{e}}(\zeta_{e})\delta(\zeta-\zeta_{e})\mathrm{e}^{i\tilde{\Delta}\zeta_{e}}+i\sqrt{\beta_{\mathrm{R}}}\sum_{j=1}^{N}\mathcal{{\tilde{E}}}^{*}\mathrm{e}^{i\tilde{\Delta}\zeta-ik_{0}z_{j}}\tilde{\phi}_{\mathrm{eS}m}^{j}(\zeta,\zeta_{e}),\label{eq:pgSm}\\
\dot{\tilde{\phi}}_{\mathrm{eS}m}^{i}(\zeta,\zeta_{e}) & =& i\sqrt{\beta_{e}}c_{\mathrm{ee}}^{im}(\zeta_{e})\delta(\zeta-\zeta_{e})\mathrm{e}^{i\tilde{\Delta}\zeta_{e}}+i\sqrt{\beta_{s}}\tilde{\phi}_{\mathrm{SS}m,m'}(\zeta,\zeta,\zeta_{e})\mathrm{e}^{-i\tilde{\Delta}\zeta}+i\sqrt{\beta_{s}}\tilde{\phi}_{\mathrm{SS}m,m'}(\zeta,\zeta_{e},\zeta)\mathrm{e}^{-i\tilde{\Delta}\zeta}\nonumber\\
&+&i\sqrt{\beta_{\mathrm{R}}}\mathcal{{\tilde{E}}}\mathrm{e}^{-i\tilde{\Delta}\zeta+ik_{0}z_{j}}\tilde{\phi}_{\mathrm{gS}m}(\zeta,\zeta_{e})+i\sqrt{\beta_{\mathrm{R}}}\tilde{\phi}_{\mathrm{RS}m}(\zeta,\zeta_{e},\zeta-\varGamma\frac{z_{i}}{v_{\mathrm{g}}})\mathrm{e}^{-i\tilde{\Delta}\zeta+ik_{0}z_{j}}\nonumber\\
&+& i\sqrt{\beta_{\mathrm{R}}}\tilde{\phi}_{\mathrm{LS}m}(\zeta,\zeta_{e},\zeta+\varGamma\frac{z_{j}}{v_{\mathrm{g}}})\mathrm{e}^{-i\tilde{\Delta}\zeta-ik_{0}z_{j}},\label{eq:peSm}\\
\eea
along with equation of motion for the two photon components corresponding to both the photons lost to the outside of the waveguide
\bea
\dot{\tilde{\phi}}_{\mathrm{SS}m,m'}(\zeta,\zeta_{e2},\zeta_{e1}) & = & i\sqrt{\beta_{s}}\tilde{\phi}^{m}_{\mathrm{eS},m'}(\zeta_{e2},\zeta_{e1})\delta(\zeta-\zeta_{e2})\mathrm{e}^{i\tilde{\Delta}\zeta_{e2}},\\
\dot{\tilde{\phi}}_{\mathrm{pS}m}(\zeta,\zeta_{e2},\zeta_{e1})& = & i\sqrt{\beta_{p}}\sum_{j=1}^{N}\tilde{\phi}^{j}_{\mathrm{eS}m}(\zeta_{e2},\zeta_{e1})\delta(\zeta-\zeta_{e2})\mathrm{e}^{i\tilde{\Delta}\zeta_{e1}-ik_{0}z_{j}},\\
\dot{\tilde{\phi}}_{\mathrm{Sp}m}(\zeta,\zeta_{e2},\zeta_{e1})& = &
i\sqrt{\beta_{\mathrm{s}}}\tilde{\phi}^{m}_{\mathrm{ep}}(\zeta_{e2},\zeta_{e1})\delta(\zeta-\zeta_{e2})\mathrm{e}^{i\tilde{\Delta}\zeta_{e1}},\label{eq:pRSm}
\eea
where the index $p = \{\text{R}, \text{L}\}$.

As before we look into the dynamics of the system in terms of the probability amplitudes after a photon goes out of the waveguide, i.e in the time-window $\zeta_{e2}+\varepsilon<\zeta<\infty$. This can be done by
formally integrating the time-evolution equations of $\phi_{\mathrm{SS}}$, $\tilde{\phi}_{\mathrm{RS}}$ and $\tilde{\phi}_{\mathrm{LS}}$,
and use the result in equation of motion of $\tilde{\phi}_{\mathrm{eS}}^{i}$. Note that the time-evolution equation for $\tilde{\phi}_{\mathrm{RS}} (\tilde{\phi}_{\mathrm{SR}})$ and $\tilde{\phi}_{\mathrm{LS}}(\tilde{\phi}_{\mathrm{SL}})$ represent two different physical processes corresponding to a state that has one photon out of the waveguide and one emitted into the waveguide, namely - either we first have decay outside the waveguide followed by emission into the waveguide or the other way around. Keeping this in mind on formally integrating the equations of motion for $\tilde{\phi}_{\mathrm{SS}m,m'}$,
$\tilde{\phi}_{\mathrm{RS}m}$ and $\tilde{\phi}_{\mathrm{LS}m}$ we get
\bea
\tilde{\phi}_{\mathrm{SS}m,m'}(\zeta_{\text{e2}}+\varepsilon,\zeta_{e2},\zeta_{e1}) & = & i\sqrt{\beta_{s}}\tilde{\phi}^{m}_{\mathrm{eS},m'}(\zeta_{e2},\zeta_{e1})\mathrm{e}^{i\tilde{\Delta}\zeta_{e2}},\\
\tilde{\phi}_{p\mathrm{S}m}(\zeta_{e2}+\varepsilon,\zeta_{e2},\zeta_{e1}) & = &  i\sqrt{\beta_{p}}\sum_{j=1}^{N}\tilde{\phi}_{\mathrm{eS}m}^{j}(\zeta_{e2},\zeta_{e1})\mathrm{e}^{i\tilde{\Delta}\zeta_{e2}-ik_{0}z_{j}},\\
\tilde{\phi}_{\mathrm{Sp}m}(\zeta_{e2}+\varepsilon,\zeta_{e2},\zeta_{e1})& = &
i\sqrt{\beta_{\mathrm{s}}}\tilde{\phi}^{m}_{\mathrm{ep}}(\zeta_{e2},\zeta_{e1})\mathrm{e}^{i\tilde{\Delta}\zeta_{e1}}
\eea
Plugging this into the equations of motion for $\tilde{\phi}_{\mathrm{gS}m}$
and $\tilde{\phi}_{\mathrm{eS}m}^{i}$, we get
\bea
\dot{\tilde{\phi}}_{\mathrm{gS}m}(\zeta,\zeta_{e}) & = & i\sqrt{\beta_{\mathrm{R}}}\sum_{j=1}^{N}\mathcal{{\tilde{E}}}^{*}\mathrm{e}^{i\tilde{\Delta}\zeta-ik_{0}z_{j}}\tilde{\phi}_{\mathrm{eS}m}^{j}(\zeta,\zeta_{e}),\\
\dot{\tilde{\phi}}^{j}_{\mathrm{eS}m}(\zeta,\zeta_{e}) & = & i\sqrt{\beta_{\mathrm{R}}}\mathcal{{\tilde{E}}}\mathrm{e}^{-i\tilde{\Delta}\zeta+ik_{0}z_{j}}\tilde{\phi}_{\mathrm{gS}m}(\zeta,\zeta_{e})-\frac{\beta_{s}}{2}\tilde{\phi}_{\mathrm{eS}m}^{j}(\zeta,\zeta_{e})-\beta_{\mathrm{R}}\sum_{l=1}^{N}\tilde{\phi}_{\mathrm{eS}m}^{l}(\zeta,\zeta_{e})\mathrm{e}^{ik_{0}(z_{j}-z_{l})}\nonumber \\
& - & \beta_{\mathrm{L}}\sum_{l=1}^{N}\tilde{\phi}_{\mathrm{eS}m}^{l}(\zeta,\zeta_{e})\mathrm{e}^{ik_{0}(z_{l}-z_{j})},\label{eq:peSm_not_final}
\eea
with the initial conditions given by
\begin{alignat}{3}
& \tilde{\phi}_{\mathrm{gS}m}(\zeta_{e}+\varepsilon,\zeta_{e}) &  & = &  & i\sqrt{\beta_{\mathrm{s}}}c_{\mathrm{e}}(\zeta_e)\mathrm{e}^{i\tilde{\Delta}\zeta_e},\nonumber\\
& \tilde{\phi}_{\mathrm{eS}m}^{j}(\zeta_{e}+\varepsilon,\zeta_{e}) &  & = &  & i\sqrt{\beta_{\mathrm{s}}}c_{\mathrm{ee}}^{j}(\zeta_e)\mathrm{e}^{i\tilde{\Delta}\zeta_e}.\label{ini_cond}
\end{alignat}
We can further simplify eq. (\ref{eq:peSm_not_final})
and get the coupled equations that govern the system dynamics for
$\zeta_{e}+\varepsilon<\zeta<\zeta_{e2}$ as
\begin{alignat}{3}
& \dot{\tilde{\phi}}_{\mathrm{gS}m}(\zeta,\zeta_{e}) &  & = &  & i\sqrt{\beta_{\mathrm{R}}}\sum_{j=1}^{N}\mathcal{{\tilde{E}}}^{*}\mathrm{e}^{i\tilde{\Delta}\zeta-ik_{0}z_{j}}\tilde{\phi}_{\mathrm{eS}m}^{i}(\zeta,\zeta_{e}),\\
& \dot{\tilde{\phi}}_{\mathrm{eS}m}^{j}(\zeta,\zeta_{e}) &  & = &  & i\sqrt{\beta_{\mathrm{R}}}\mathcal{{\tilde{E}}}\mathrm{e}^{-i\tilde{\Delta}\zeta+ik_{0}z_{i}}\tilde{\phi}_{\mathrm{gS}m}(\zeta,\zeta_{e})-\frac{1}{2}\tilde{\phi}_{\mathrm{eS}m}^{j}(\zeta,\zeta_{e})-\beta_{\mathrm{R}}\sum_{j<l}^{N}\tilde{\phi}_{\mathrm{eS}m}^{l}(\zeta,\zeta_{e})\mathrm{e}^{ik_{0}(z_{i}-z_{j})}\nonumber \\
&  &  & - &  &\beta_{\mathrm{L}}\sum_{j>l}^{N}\tilde{\phi}_{\mathrm{eS}m}^{j}(\zeta,\zeta_{e})\mathrm{e}^{ik_{0}(z_{j}-z_{i})}.\label{eq:peSm_final}
\end{alignat}
where we have used $\beta_{\text{R}}+\beta_{\text{L}}+\beta_{\text{s}} = 1$.The tilde in the above equations represent the quantities as function of the dimensionless time variable. Keeping this in mind we will drop the tilde for all further discussion. 

We next define two time windows $0<\zeta<\zeta_{e}+\varepsilon$ and $\zeta_{e}+\varepsilon<\zeta<\infty$, to investigate the dynamics of the photon scattered from the emitters. We see that in these two time domains the above set of Eqs. (\ref{a4})-(\ref{a6}) and (\ref{a7})-(\ref{a9}) becomes uncoupled. For $0<\zeta<\zeta_{e}+\varepsilon$ we can formally integrate the equations for $\phi_{g\text{R}}$, $\phi_{g\text{L}}$, $\phi_{e\text{R}}^{j}$ and $\phi_{e\text{L}}^{j}$ as
\bea
\label{a14}
 \tilde{\phi}_{g\text{R}}(\zeta_{e}+\varepsilon,\zeta_{e}) & = & i\sqrt{\beta_{\text{R}}}\sum_{j=1}^{N}c_{e}^{j}(\zeta_e)e^{i\Delta\zeta_{e}-ik_{0}\text{z}_{j}},\\
 \label{a15}
\tilde{\phi}_{g\text{L}}(\zeta_e+\varepsilon,\zeta_{e}) & = & i\sqrt{\beta_{\text{L}}}\sum_{j=1}^{N}c_{e}^{j}(\zeta_{e})e^{i\Delta\zeta_{e}+ik_{0}\text{z}_{j}},\\
 \label{a16}
 \tilde{\phi}_{e\text{R}}^{j}(\zeta_e+\varepsilon,\zeta_{e}) & = & i\sqrt{\beta_{\text{R}}}\sum_{l<j}^{N}c_{ee}^{lj}(\zeta_e)e^{i\Delta\zeta-ik_{0}\text{z}_{l}}+i\sqrt{\beta_{\text{R}}}\sum_{l>j}^{N}c_{ee}^{jl}(\zeta)\mathrm{e}^{i\Delta\zeta-ik_{0}\text{z}_{l}},\\
 \label{a17}
\tilde{\phi}_{e\text{L}}^{j}(\zeta_e+\varepsilon,\zeta_{e}) & = &i\sqrt{\beta_{\text{L}}}\sum_{l<j}^{N}c_{ee}^{lj}(\zeta_e)e^{i\Delta\zeta+ik_{0}\text{z}_{l}}+  i\sqrt{\beta_{\text{L}}}\sum_{l>j}^{N}c_{ee}^{jl}(\zeta_e)e^{i\Delta\zeta+ik_{0}\text{z}_{l}}.
\eea

Using the above set of solutions, Eqs. (\ref{a14})-(\ref{a17}) and substituting for $\phi_{g\text{R}}$, $\phi_{g\text{L}}$, $\phi_{e\text{R}}^{j}$, $\phi_{e\text{L}}^{j}$  and $\phi^{j}_{e\text{S}m}, \phi_{g\text{S}m}$ from Eq. (\ref{ini_cond}) into the equations of motion for $c_{e}^{j}$ and $c_{ee}^{jl}$ we get the equations of motion for the amplitude of the emitter's state in the time window $0<\zeta<\zeta_{\mathrm{e}}+\varepsilon$
as 
\bea
\label{a20} 
\dot{c}_{g}(\zeta) & = &i\sqrt{\beta_{\text{R}}}\sum_{i=1}^{N}\mathcal{{E}}^{*}e^{i\Delta\zeta-ik_{0}\text{z}_{j}}c_{e}^{j}(\zeta),\\
\label{a21}
\dot{c}_{e}^{j}(\zeta) & = & i\sqrt{\beta_{\text{R}}}\mathcal{{E}}e^{-i\Delta\zeta+ik_{0}\text{z}_{j}}c_{g}(\zeta)-\frac{1}{2}c_{e}^{j}(\zeta)+i\sqrt{\beta_{\text{R}}}\sum_{l<j}\mathcal{{E}}^{*}e^{i\Delta\zeta-ik_{0}\text{z}_{l}}c_{ee}^{lj}(\zeta)+i\sqrt{\beta_{\text{R}}}\sum_{j<l}\mathcal{{E}}^{*}e^{i\Delta\zeta-ik_{0}\text{z}_{l}}c_{ee}^{jl}(\zeta)\nonumber\\
& - &\beta_{\text{R}}\sum_{l<j}c_{e}^{j}(\zeta)e^{ik_{0}(\text{z}_{j}-\text{z}_{l})}-\beta_{\text{L}}\sum_{l>j}c_{e}^{l}(\zeta)e^{ik_{0}(\text{z}_{l}-\text{z}_{j})},\\
\label{a22}
 \dot{c}_{ee}^{jl}(\zeta) & = & i\sqrt{\beta_{\text{R}}}\mathcal{{E}}e^{-i\Delta\zeta+ik_{0}\text{z}_{l}}c_{e}^{j}(\zeta)+i\sqrt{\beta_{\text{R}}}\mathcal{{E}}e^{-i\Delta\zeta+ik_{0}\text{z}_{j}}c_{e}^{l}(\zeta)-c_{ee}^{jl}(\zeta)-\beta_{\text{R}}\sum_{l'<j}c_{ee}^{l'j}(\zeta)e^{ik_{0}(\text{z}_{l}-\text{z}_{l'})}\nonumber\\
 &-&\beta_{\text{R}}\sum_{l'<l,l'>j}c_{ee}^{jl'}(\zeta)e^{ik_{0}(\text{z}_{l}-\text{z}_{l'})}- \beta_{\text{R}}\sum_{j'<j,}c_{ee}^{j'l}(\zeta)e^{ik_{0}(\text{z}_{j}-\text{z}_{j'})}-\beta_{\text{L}}\sum_{l'>l}c_{ee}^{jl'}(\zeta)e^{ik_{0}(\text{z}_{l'}-\text{z}_{l})}\nonumber\\
&- &\beta_{\text{L}}\sum_{j'>j,j'<l}^{N}c_{ee}^{j'l}(\zeta)e^{ik_{0}(\text{z}_{j'}-\text{z}_{j})}-\beta_{\text{L}}\sum_{j'>l}^{N}c_{ee}^{lj'}(\zeta)e^{ik_{0}(\text{z}_{j'}-\text{z}_{j})}.\nonumber\\
\eea
where as before we have used $\beta_{\text{R}}+\beta_{\text{L}}+\beta_{s} = 1$. 

\section{Detail calculation of the first order and second order correlation function}
The standard definition of normal ordered first and second order correlation functions are given as
\bea
\label{eqc1}
\mathbf{G}^{(1)} & = &\langle : E^{\dagger}(t)E(t): \rangle\\
\mathbf{G}^{(2)} & = &\langle : E^{\dagger}(t)E^{\dagger}(t+\tau)E(t+\tau)E(t) :\rangle.
\eea
For our work the above equation translates to Eq. (\ref{eq19}) and Eq. (\ref{eq20}) in the main text as
\bea
\label{eqcc2}
\text{G}^{(1)}(t_d) & = & v_{g}\langle\tilde{\Psi}(T)|\left(\hat{E}^{\dagger}_{\text{R}}(\text{z}_{t_d})+\mathcal{E}^\ast\right)\left(\hat{E}_{\text{R}}(\text{z}_{t_d})+\mathcal{E}\right)\Ket{\tilde{\Psi}(T)},\\
\label{eqc3}
\text{G}^{(2)}(t_d+\tau_d,t_d) & = & v^{2}_{g}\Bra{\tilde{\Psi}(T)}\left(\hat{E}^\dagger_{\text{R}}(\text{z}_{t_d+\tau_d})+\mathcal{E}^\ast\right)\left(\hat{E}^\dagger_{\text{R}}(\text{z}_{t_d})+\mathcal{E}^\ast\right) \left(\hat{E}_{\text{R}}(\text{z}_{t_d+\tau_d})+\mathcal{{E}}\right)\left(\hat{E}_{\text{R}}(\text{z}_{t_d})+\mathcal{{E}}\right)\Ket{\tilde{\Psi}(T)},
\eea
where $|\tilde{\Psi}(T)\rangle$ is the $N$-emitter wave-function ansatz given by Eq. (\ref{eq8}) in the main text. We evaluated Eq. (\ref{eqc2}) and Eq. (\ref{eqc3}) in the main text under the approximation of a weak  excitation field $|\mathcal{E}| \ll 1$. Hence we kept only the leading order terms in $\mathcal{E}$ in the derived expressions of Eq. (\ref{eqcc2}) and Eq. (\ref{eqc3}). Here in this appendix we provide the complete expression for the first and second order correlation functions in all orders of $\mathcal{E}$, consistent with the truncation of the wave-function ansatz to two-photon emissions. 

The first order correlation function $\text{G}^{(1)}(\zeta_e)$ can be written as a sum of four expectation values in the form: 
\bea
\label{eqc4}
\text{G}^{(1)}(t_d) & = & v_{g}\left(\langle\hat{E}^{\dagger}_{\text{R}}(\text{z}_{\zeta_d})\hat{E}_{\text{R}}(\text{z}_{\zeta_d})\rangle+\langle\mathcal{E}^\ast\hat{E}_{\text{R}}(\text{z}_{\zeta_d})\rangle+\langle\hat{E}^\dagger_{\text{R}}(\text{z}_{\zeta_d})\mathcal{E}\rangle+|\mathcal{E}|^{2}\right),
\eea
with the individual terms defined as 
\bea
\label{eqc5}
\langle\hat{E}^{\dagger}_{\text{R}}(\text{z}_{\zeta_d})\hat{E}_{\text{R}}(\text{z}_{\zeta_d})\rangle & = & \left|\tilde{\phi}_{g\mathrm{R}}(\zeta_{T},\zeta_{e})\right|^{2}+\sum_{i=1}^{N}\left|\tilde{\phi}_{e\mathrm{R}}^{i}(\zeta_{T},\zeta_{e})\right|^{2}+\int\mathrm{d}\zeta'_{e}\left|\tilde{\phi}_{\mathrm{RR}}(\zeta_{T},\zeta_{e},\zeta'_{e})\right|^{2}+\int\mathrm{d}\zeta'_{e}\left|\tilde{\phi}_{\mathrm{RR}}(\zeta_{T},\zeta'_{e},\zeta{}_{e})\right|^{2}\nonumber \\
& + & \int\mathrm{d}\zeta'_{e}\left|\tilde{\phi}_{\mathrm{RL}}(\zeta_{T},\zeta_{e},\zeta'_{e})\right|^{2}+ \int\mathrm{d}\zeta'_{e}\left|\tilde{\phi}_{\mathrm{LR}}(\zeta_{T},\zeta'_{e},\zeta_{e})\right|^{2}+\int\mathrm{d}\zeta_{e}\sum_{m=1}^{N}\left|\tilde{\phi}_{\mathrm{RS},m}(\zeta_{T},\zeta'_{e},\zeta_{e})\right|^{2},\nonumber\\
& + & \int\mathrm{d}\zeta'_{e}\sum_{m=1}^{N}\left|\tilde{\phi}_{\mathrm{SR},m}(\zeta_{T},\zeta'_{e},\zeta_{e})\right|^{2}\\
\label{eqc6}
\langle\hat{E}^\dagger_{\text{R}}(\text{z}_{\zeta_d})\mathcal{E}\rangle & =&
\tilde{\phi}_{g\mathrm{R}}(\zeta_{T},\zeta_{e})\mathcal{{\tilde{E}}}c_{g}(\zeta_{T})+\sum_{i=1}^{N}\tilde{\phi}_{e\mathrm{R}}^{i}(\zeta_{T},\zeta_{e})\mathcal{{\tilde{E}}}c_{e}^{i}(\zeta_{T})\int\mathrm{d}\zeta'_{e}\tilde{\phi}_{\mathrm{RR}}(\zeta_{T},\zeta_{e},\zeta'_{e})\mathcal{{\tilde{E}}}\tilde{\phi}_{g\mathrm{R}}(\zeta_{T},\zeta'_{e})\nonumber \\
& + & \int\mathrm{d}\zeta'_{e}\tilde{\phi}_{\mathrm{RR}}(\zeta_{T},\zeta'_{e},\zeta_{e})\mathcal{{\tilde{E}}}\tilde{\phi}_{g\mathrm{R}}(\zeta_{T},\zeta'_{e})+ \int\mathrm{d}\zeta'_{e}\tilde{\phi}_{\mathrm{RL}}(\zeta_{T},\zeta_{e},\zeta'_{e})\mathcal{{\tilde{E}}}\tilde{\phi}_{g\mathrm{L}}(\zeta_{T},\zeta'_{e})\nonumber\\
& + &\int\mathrm{d}\zeta'_{e}\tilde{\phi}_{\mathrm{LR}}(\zeta_{T},\zeta'_{e},\zeta_{e})\mathcal{{\tilde{E}}}\tilde{\phi}_{g\mathrm{L}}(\zeta_{T},\zeta'_{e})+ \int\mathrm{d}\zeta_{e}\sum_{m=1}^{N}\tilde{\phi}_{\mathrm{RS},m}(\zeta_{T},\zeta'_{e},\zeta_{e})\mathcal{{\tilde{E}}}\tilde{\phi}_{g\mathrm{S},m}(\zeta_{T},\zeta_{e})\nonumber\\
& + & \int\mathrm{d}\zeta'_{e}\sum_{m=1}^{N}\tilde{\phi}_{\mathrm{SR},m}(\zeta_{T},\zeta'_{e},\zeta_{e})\mathcal{{\tilde{E}}}\tilde{\phi}_{g\mathrm{S},m}(\zeta_{T},\zeta'_{e}),\\
\label{eqc7}
 \langle\mathcal{E}^\ast\hat{E}_{\text{R}}(\text{z}_{\zeta_d})\rangle& = & \langle(\hat{E}^\dagger_{\text{R}}(\text{z}_{\zeta_d})\mathcal{E})^\dagger\rangle
\eea

The second order correlation function $\text{G}^{(2)}$ defined above can be evaluated in the form 
\bea
\label{eqc8}
\text{G}^{(2)}(t_d+\tau_d,t_d) & = & v^{2}_{g}\left| \left(\hat{E}_{\text{R}}(\text{z}_{t_d+\tau_d})+\mathcal{{E}}\right)\left(\hat{E}_{\text{R}}(\text{z}_{t_d})+\mathcal{{E}}\right)|\tilde{\Psi}(T)\rangle\right|^{2}\\
\label{eqc9}
\text{G}^{(2)}(\zeta'_{e}, \zeta_{e}) & = &v^{2}_{g}\bigg|\tilde{\phi}_{\text{RR}}(\zeta_{T}, \zeta'_{e}, \zeta_{e})+\tilde{\phi}_{\text{RR}}(\zeta_{T}, \zeta_{e}, \zeta'_{e}) +\tilde{\mathcal{E}}(\zeta'_{e})\tilde{\phi}_{g\text{R}}(\zeta_{T},\zeta_{e})+\tilde{\phi}_{g\text{R}}(\zeta_{T},\zeta'_{e})\tilde{\mathcal{E}}(\zeta_{e})+\sum_{i=1}^{N}\tilde{\phi}^{i}_{e\text{R}}(\zeta_{T},\zeta'_{e})\tilde{\mathcal{E}}(\zeta_{e})\nonumber\\
& + &\tilde{\mathcal{E}}(\zeta_{e})\sum_{i=1}^{N}\tilde{\phi}^{i}_{e\text{R}}(\zeta_{T},\zeta'_{e})+\tilde{\mathcal{E}}(\zeta'_e)\tilde{\phi}_{\text{RL}}(\zeta_T, \zeta'_{e}, \zeta_{e}) + \tilde{\mathcal{E}}(\zeta_e)\tilde{\phi}_{\text{LR}}(\zeta_T, \zeta_{e}, \zeta'_{e})\nonumber\\ 
&+ &\int\mathrm{d}\zeta'_{e}\sum_{m=1}^{N}\tilde{\phi}_{\mathrm{RS},m}(\zeta_{T},\zeta_{e},\zeta'_{e})\tilde{\mathcal{E}}(\zeta_{e})+ \int\mathrm{d}\zeta_{e}\sum_{m=1}^{N}\tilde{\phi}_{\mathrm{SR},m}(\zeta_{T},\zeta'_{e},\zeta_{e})\tilde{\mathcal{E}}(\zeta'_{e})+\tilde{\mathcal{E}}(\zeta'_{e})\tilde{\mathcal{E}}(\zeta_e)\bigg|^{2}
\eea
\end{widetext}

\end{document}